\def\la{$_<\atop{^\sim}$}
\def\ga{$_>\atop{^\sim}$}

\def\cm2{cm$^{-2}$}

\def\etal{et~al.}
\def\e{et~al.}

\def\c2{C~{\sc ii}}
\def\c4{C~{\sc iv}}
\def\fe2{Fe~{\sc ii}}
\def\fe3{Fe~{\sc iii}}
\def\mg1{Mg~{\sc i}}
\def\mg2{Mg~{\sc ii}}
\def\si2{Si~{\sc ii}}
\def\si4{Si~{\sc iv}}
\def\al2{Al~{\sc ii}}
\def\al3{Al~{\sc iii}}
\def\o1{O~{\sc i}}
\def\n1{N~{\sc i}}
\def\h1{H~{\sc i}}

\def\approxlt{\mathrel{\spose{\lower 3pt\hbox{$\sim$}}
        \raise 2.0pt\hbox{$<$}}}
\def\approxgt{\mathrel{\spose{\lower 3pt\hbox{$\sim$}}
        \raise 2.0pt\hbox{$>$}}}

\documentstyle[psfig,rotating]{mn}

\input{psfig.sty}

\newif\ifAMStwofonts

\title[Statistical, Kinematic and Chemical Properties of Sub-DLAs]{A Homogeneous Sample of Sub-DLAs II: Statistical, Kinematic and Chemical Properties}

\author[C\'eline P\'eroux et al.]
        {C\'eline P\'eroux$^{1,2}$\thanks{Marie Curie
        Fellow}\thanks{e-mail: peroux@ts.astro.it}, Miroslava
        Dessauges-Zavadsky$^{3,4}$, Sandro D'Odorico$^{3}$,
\newauthor
Tae Sun Kim$^{3,2}$ \& Richard G. McMahon$^{2}$\\
$1$ Osservatorio Astronomico di Trieste, Via G. B. Tiepolo 11, 34131 Trieste, Italy.\\
$2$ Institute of Astronomy, Madingley Road, Cambridge CB3 0HA, UK.\\
$3$ European Southern Observatory, Karl-Schwarzchild-Str. 2, 85748 Garching bei M\"unchen, Germany.\\
$4$ Observatoire de Gen\`eve, 1290 Sauverny, Switzerland.\\
}

\date{}

\pagerange{\pageref{firstpage}--\pageref{lastpage}}

\begin{document}

\maketitle

\label{firstpage}

\begin{abstract}

Damped Ly$\alpha$ Systems (DLAs), with N(HI) $> 2 \times 10^{20}$
cm$^{-2}$, observed in the spectra of quasars have allowed us to
quantify the chemical content of the Universe over cosmological
scales. Such studies can be extended to lower column densities, in the
sub-DLA range ($10^{19}$ $<$ N(HI) $<$ 2 $\times 10^{20}$ cm$^{-2}$),
which are systems believed to contain a large fraction of the neutral
hydrogen at $z>3.5$. In this paper, we use a homogeneous sample of
sub-DLAs from the ESO UVES archives presented in Paper~I
(Dessauges-Zavadsky \e\ 2003), to observationally determine for the
first time the shape of the column density distribution, $f(N)$, down
to N(HI) $=10^{19}$ cm$^{-2}$. The results are in good agreement with
the predictions from P\'eroux \e\ (2003). We also present the
kinematic and clustering properties of this survey of sub-DLAs which
appear to be marginally different from the DLAs. We compare low- and
high-ionization transition widths and find that the sub-DLAs
properties roughly span the parameter space of DLAs. We also find
hints of an increase of metallicity in systems with larger velocity
widths in the metal lines, although the statistical significance of
this result is low.

We then analyse the chemical content of this sample in conjunction
with a compilation of abundances from 72 DLAs taken from the
literature. As previously reported, the individual metallicities
traced by [Fe/H] of these systems evolve mildly with
redshift. Moreover, we analyse the HI column density-weighted mean
abundance which is believed to be an indicator of the Universe's
metallicity. Although the number statistics is limited in the current
sample, the results suggest a slightly stronger evolution of this
quantity in the sub-DLA range. The effect is predominant at $z<2$ and
most of the evolution observed lies in this redshift
range. Observational arguments support the hypothesis that the
evolution we probe in the sub-DLA range is {\it not} due to their
lower dust content. Therefore these systems might be associated with a
different class of objects which better trace the overall chemical
evolution of the Universe. Finally, we present abundance ratios of
[Si/Fe], [O/Fe], [C/Fe] and [Al/Fe] for sub-DLAs in conjunction with
DLA measurements from the literature. The elemental ratios in sub-DLAs
are comparable with those from DLAs. It is difficult to decipher
whether the observed values are the effect of nucleosynthesis or are
due to differential dust depletion. The metallicities are compared
with two different sets of models of galaxy evolution in order to
provide constraints on the morphology of quasar absorbers.
\end{abstract}

\begin{keywords}
galaxies: abundance -- galaxies: high-redshift -- quasars: absorption
lines -- quasars:
\end{keywords}

\section{Introduction}

In addition to traditional emission studies, absorption systems along
the line-of-sight to distant quasars provide a completely independent
probe of galaxy evolution. The highest column density Damped
Ly$\alpha$ systems (hereafter DLAs) have N(HI) $> 2 \times 10^{20}$
atoms cm$^{-2}$. The reason why damped systems are a cosmologically
important population is that they contain most of the neutral gas in
the Universe at $z>1$ (Lanzetta et al.\ 1991; Wolfe \e\ 1995;
Storrie-Lombardi et al.\ 1996b). Since the metal content of these
systems can be determined with rather high precision up to high
redshift, they are a powerful tool to study the chemical evolution of
galaxies. In particular, a way to trace the metallicity of the
Universe is provided by estimating the ratio of the total metal
content to the total gas content measured in these systems (Pettini
\e\ 1997; Pei, Fall \& Hauser 1999). Using such techniques, Prochaska
\& Wolfe (2002), in agreement with previous work, find that the 
hydrogen column density-weighted [Fe/H] metallicities have similar
values from $1.5<z<3.5$. The results are in disagreement with
expectations since models of cosmic chemical evolution (e.g. Pei \&
Fall 1995) predict evolution of global metallicity with cosmic
time. Savaglio (2000), however, has used 50 DLAs and lower column
density systems and finds clear evidence for redshift evolution,
although this sample suffers high inhomogeneity.

In a recent study, P\'eroux \e\ (2001) have shown that sub-DLAs,
defined as systems with $10^{19}<$ N(HI) $< 2 \times 10^{20}$ atoms
cm$^{-2}$, play a major role, especially at higher
redshifts. The authors postulate that at $z>3.5$, 45\% of the neutral
gas mass is in sub-DLAs. These predictions are based on a constrained
extrapolation of the quasar absorber column density distribution,
$f(N)$, to lower column densities assuming the distribution can be
fitted by a gamma function.
This suggests that the metallicities of sub-DLAs require detailed
study in order to obtain a complete picture of the
redshift evolution of the metallicity of the Universe. For this purpose, we
have constructed a homogeneous sample of sub-DLAs based on
high-resolution quasar spectra from the ESO UVES/VLT archives
(Dessauges-Zavadsky \e\ 2003, hereafter Paper I). Sub-DLAs can be
easily picked up since at these column densities damping wings are
already formed and the rest equivalent width is $W_{\rm rest}>2.5$
\AA.

In this paper, we analyse the sample of sub-DLAs presented in Paper
I. We first use these data to establish the column density
distribution in a range unprobed before. These results have direct
implications for the determination of the neutral gas content of
quasar absorbers and its evolution with redshift. We also emphasize
the limitations of the present sample in terms of redshift path
surveyed. The clustering properties of our sample of sub-DLAs are
described in section 3. In section 4, we discuss the kinematic
properties of the low- and high-ionization transitions associated with
the sub-DLAs presented in Paper I. These results can be used to
directly constrain some of the semi-analytic models recently
presented in the literature. In section 5, we present the results of
chemical abundance determinations in sub-DLAs in conjunction with the
abundances in DLAs taken from the literature and provide information on
the global metallicity evolution of both individual quasar absorbers
and the Universe. We also cross-correlate these properties with the
kinematic information we present in previous
sections. Relative abundances are then analysed and compared with
recent models of galaxy formation and evolution, putting further
constraints on the morphology of quasar absorbers. 

\begin{table*}\centering
\begin{minipage}{140mm}
\caption{$\Delta z$ covered by each quasar used from the UVES archives and the associated sub-DLA systems.}
\begin{tabular}{lcllll}

\hline
Quasar &$z_{\rm em}$ &Ly-$\alpha$ forest coverage &$z_{\rm DLA}$
&$z_{\rm sub-DLA}$ &sub-DLA\\ Name & &($z_{\rm min}$ - $z_{\rm max}$)$^a$ & &
&$\rm N(HI)$ \\
\hline
Q0000$-$2620     &4.11	&2.533--3.365 $\&$ 3.423--4.105   &3.390		&...	&...	\\
BR J0307$-$4945  &4.75	&3.119--3.903 $\&$ 4.034--4.455   &4.466		&...    &...   	\\
...              &...   & $\&$ 4.482--4.774               &...			&...	&...\\
Q0347$-$3819     &3.23 	&2.017--3.014                     &3.025	     	&...    &...	\\
Q0841$+$129      &2.50  &1.903--2.362 $\&$ 2.387--2.466   &2.375	        &...	&...	\\
...              &...   & $\&$ 2.484--2.497               &2.476		&...	&...\\
HE 0940$-$1050   &3.05  &1.719--2.183 $\&$ 2.933--3.046   &...			&...    &...	\\
Q1038$-$272      &2.32  &2.107--2.357                     &...			&...	&...	\\
Q1101$-$264$^b$  &2.14	&1.519--2.117                     &...		        &1.838$^b$&19.50$\pm$0.05\\
HE 1104$-$1805$^c$&2.31 &1.609--1.654 $\&$ 1.670--2.307   &1.662		&...	&...	\\
Q1151$+$068      &2.76  &1.797--2.756                     &1.774$^d$		&...	&...	\\  
PKS 1157$+$014   &1.99  &1.559--1.918 $\&$ 1.982--1.987   &1.44			&...	&...	\\
Q1223$+$1753     &2.94 	&2.044--2.446 $\&$ 2.484--2.936   &2.466		&2.557	&19.32$\pm$0.15	\\
...              &...	&...                              &...			&...	&...	\\
PKS 1232$+$0815  &2.57 	&1.704--2.327 $\&$ 2.350--2.566   &2.338		&...	&...	\\ 
Q1409$+$095      &2.86  &2.126--2.448 $\&$ 2.463--2.856   &2.456		&2.668	&19.75$\pm$0.10\\	
Q1444$+$014      &2.21  &1.710--2.082 $\&$ 2.091--2.207   &...			&2.087	&20.18$\pm$0.10	\\
Q1451$+$123      &3.25  &2.168--2.462 $\&$ 2.475--2.995   &2.469	   	&3.171$^e$&19.70$\pm$0.15\\
...              &...   &...                              &2.255		&...	&...	\\
Q1511$+$090      &2.88 	&1.720--2.876                     &...			&2.088	&19.47$\pm$0.10	\\
Q2059$-$360      &3.09  &2.461--3.068                     &3.083		&2.507	&20.21$\pm$0.10	\\
Q2116$-$358      &2.34  &1.710--1.992 $\&$ 2.000--2.337   &...			&1.996	&20.06$\pm$0.10	\\
Q2132$-$433      &2.42  &1.710--1.907 $\&$ 1.923--2.417   &1.914		&...	&...	\\
PSS J2155$+$1358 &4.26 	&2.913--3.300 $\&$ 3.337--3.606   &3.316		&3.142  	&19.94$\pm$0.10	\\
...              &...	& $\&$ 3.689--4.255               &...			&3.565  	&19.37$\pm$0.15\\
...              &...	&...                              &...			&4.212	&19.61$\pm$0.15\\
Q2206$-$199      &2.56  &1.720--1.914 $\&$ 1.917--2.071   &1.920    		&...	&...	\\
...              &...   & $\&$ 2.083--2.556               &2.076		&...	&...\\
PSS J2344$+$0342 &4.24  &2.889--3.201 $\&$ 3.241--3.606   &3.220	    	&3.882  	&19.50$\pm$0.10\\
...              &...   &$\&$ 3.689--4.295                &...			&...	&...\\
\hline		       
\end{tabular}

\vspace{0.5cm}

$^a$ $z_{\rm min}$ corresponds to the point where the signal-to-noise
ratio was too poor to find absorption features and $z_{\rm max}$ is
3000 km s$^{-1}$ bluewards of the Ly-$\alpha$ emission of the quasar. The
holes in the redshift intervals correspond to either a DLA or a gap in
the spectrum due to non-overlapping settings.\\

$^b$ This quasar has been observed as part as the Science Verification
of UVES for the study of the Ly$\alpha$ forest. As such, it does
{\it not} fulfill our criteria for its redshift path to be included
in our ``statistical sample''. Nevertheless, an analysis of the
spectrum reveals the presence of a sub-DLA system at
$z_{abs}=1.838$ and we undertook the determination of the abundance of
this absorber.

$^c$ This quasar is gravitationally lensed. Only the brightest line
of sight was included in our study.\\

$^d$ This system is situated outside our spectral coverage.\\

$^e$ Our spectrum does {\it not} cover the Ly$\alpha$ line of this
sub-DLA ($z_{\rm abs} = 3.171$, $\lambda_{\rm obs} \sim 5070$ \AA). It
has been identified thanks to the HI column density measurement
reported by Bechtold (1994) and Petitjean, Srianand \& Ledoux
(2000). Several lines from the Lyman series and metals are available,
hence, we undertook the determination of the metal content of this
system but did {\it not} include it in our ``statistical sample''.\\

\end{minipage}
\label{t:sub-DLA_stat}
\end{table*}

\section{Statistical Properties of sub-DLAs}
\subsection{Sample Definition}

The sample of quasars used for this study is presented in Paper I and
their characteristics are summarised in
Table~\ref{t:sub-DLA_stat}. This lead to a set of 22 quasars studied in
various observational programmes. In order to build the
``statistical sample'', we have ignored the quasars which have been
targeted for the study of the Ly$\alpha$ forest (for the
analysis of voids, low column density column density distribution,
etc). Indeed these have been {\it pre-selected} for not having a LLS
in their spectrum, precisely the type of feature we are looking for in
the present study. As reported in Paper I, a sub-DLA at $z_{abs}=1.838$ 
was found in Q1101$-$264, a quasar observed during UVES Science Verification. 
This absorber is included
in the abundance analysis, although for consistency, neither the
redshift path of the quasar nor the sub-DLA are used in the
``statistical sample''. This selection process resulted in a sample of
21 quasars suitable for studying the statistical properties of
sub-DLAs. 

Similarly, sub-DLAs which do not have their Ly$\alpha$ feature within
the spectral coverage are automatically excluded, although the
determination of HI column density might be possible from other lines
of the Lyman series. This is the case of the sub-DLA at $z_{\rm abs}
=3.171$ towards Q1451$+$123, for which we undertook a detailed
abundance study similar to the other systems, but which is not
included in our ``statistical sample''. The resulting sample of
sub-DLAs is thus composed of 10 systems. Although we looked for any
system with N(HI) $>10^{19}$ atoms cm$^{-2}$, the smallest column
density sub-DLA detected in this sample is N(HI) $=10^{19.32}$ atoms
cm$^{-2}$.

The large majority of the quasars in the sample were observed with
UVES because there was a DLA along their line-of-sight, as can be
attested from Table~\ref{t:sub-DLA_stat}. There is no {\it a priori}
reason that this creates a bias for our statistical analysis.

Table~\ref{t:sub-DLA_stat} lists the minimum ($z_{\rm min}$) and
maximum ($z_{\rm max}$) redshifts along which a sub-DLA could be
detected along each quasar line-of-sight. $z_{\rm min}$ corresponds to
the point where the signal-to-noise ratio was too low to find
absorption features at the sub-DLA threshold of $W_{\rm rest}=2.5$ \AA,
and $z_{\rm max}$ is 3000 km s$^{-1}$ blueward of
the Ly$\alpha$ emission of the quasar. We took care to exclude the
DLA regions and the gaps in the spectrum due to non-overlapping settings
when computing the redshift path surveyed.

\subsection{Number Density}

The number density of quasar absorbers is the number of absorbers,
$n$, per unit redshift $dz$, i.e., $dn/dz = n(z)$. $dz$ is computed by
summing up the redshift paths surveyed along the line-of-sight of each
of the quasar studied as given in Table~\ref{t:sub-DLA_stat}. 
$n(z)$ is a directly observable quantity, although, its
interpretation is dependent on the geometry of the Universe. Indeed,
the evolution of the number density of absorbers with redshift is the
intrinsic evolution of the true number of absorbers combined with
effects due to the expansion of the Universe.

Measuring the incidence of sub-DLAs down to N(HI) $=10^{19}$ atoms
cm$^{-2}$ in a given sample of quasars requires a number of high
resolution spectra in order to unambiguously select all the absorption
systems. Nevertheless, indirect information is also provided by the
number of LLS in a given quasar sample. P\'eroux \etal\ (2003) fitted
the observed cumulative number of absorbers of both DLAs and LLSs with
a $\Gamma$-distribution as suggested by Pei \& Fall (1995) and first
implemented by Storrie-Lombardi, Irwin \& McMahon (1996a). This puts
constraints on the number of systems in the $10^{19} <$ N(HI) $<
2\times 10^{20}$ atoms cm$^{-2}$ column density range. They then used
these predictions to compute the expected number density of
sub-DLAs. The results from this work are tabulated in the last column
of 
Table 2 for comparison with the observations
presented here.

Excluding Q1101$-$264, we have sampled a
total redshift path $dz=17.5$ and find a total of 10 sub-DLAs. The
comparison with the predictions of P\'eroux \etal\ (2003) show a good
agreement in the intermediate redshift range. At
$z>3.5$, the low number statistics makes the results more
uncertain. Nevertheless, it is important to note that in the quasar
sample studied here all but 4 of the objects have $z_{\rm
em}<3.5$. The sub-DLAs are thus predominantly located in the
intermediate redshift range $2<z_{\rm abs}<3.5$. Indeed, we find 7
sub-DLAs at $z_{\rm abs}<3.5$, where we have surveyed $dz=14.4$, while
we find 3 sub-DLAs at $z_{\rm abs}
\ga 3.5$, where we have only surveyed $dz=3.1$. Consequently, the
determination of $n(z)$ at $z_{\rm abs}\ga 3.5$ is not possible with
the current set of data. The number of sub-DLAs and associated
redshift paths are summarised in 
Table 2 for various redshift ranges.

\begin{table}
\begin{center}
\caption{This table gives the redshift path surveyed, $dz$, and 
corresponding physical distance interval, $dX$ (in a $\Omega_M=0.3$,
$\Omega_{\Lambda}=0.7$ cosmology) for various redshift ranges. The
observed number density of sub-DLAs is from our ``statistical sample''
of 21 quasars from the UVES archive spectra. The predicted number
density is from P\'eroux \etal\ (2003) who used a fit to the observed
number of DLAs with an additional constraint from the total number of
LLS.}
\begin{tabular}{lccccc}
\hline
z range &$dz$ &$dX$ &number &$n(z)$  &$n(z)$\\ 
&obs &obs &of sub-DLAs &obs &       predicted\\
\hline
$<$3.5   &14.4        &47.0       &7            &0.49	   &0.46\\
$>$3.5   &3.1	      &12.5       &3            &0.97	   &1.66\\
\hline						 
1.5-2.5  &8.7         &27.0       &3            &0.34      &0.39\\
2.5-3.0  &3.8         &13.0       &3            &0.79      &0.46\\
3.0-3.5  &1.9         &7.0        &1            &0.53      &0.67\\
3.5-4.0  &1.7         &6.8        &2            &1.18      &1.52\\
4.0-5.0  &1.4         &5.7        &1            &0.71      &2.00\\
total    &17.5        &59.5       &10           &0.57      &0.58\\
\hline
\end{tabular}
\end{center}
\label{t:nz}
\end{table}

\subsection{Column Density Distribution}

\subsubsection{Observationally Constraining $f(N)$ down to log $ N(HI) = 19.0$}

\begin{figure}
\psfig{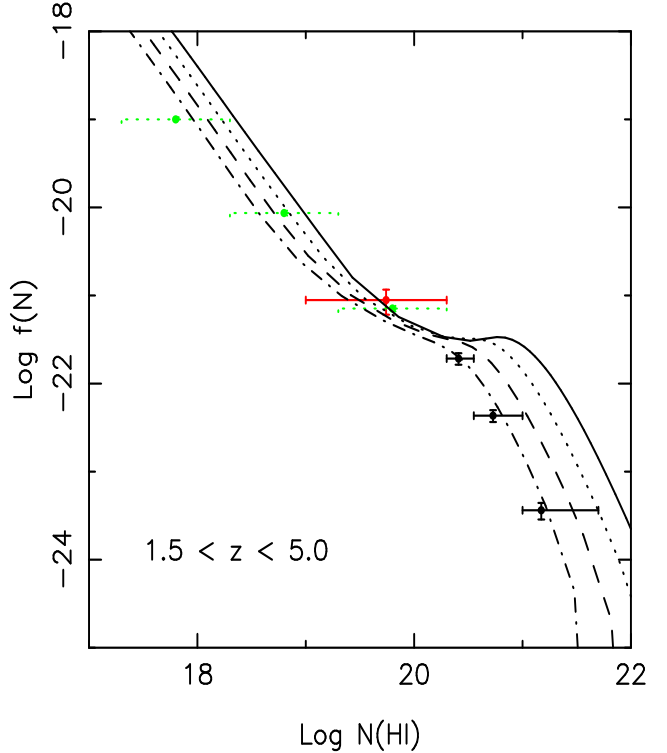}
\caption{The solid bins at log $ N(HI) > 20.3$ atoms 
cm$^{-2}$ are measurements of the column density distribution
of DLAs from P\'eroux \etal\ (2002). The dashed bins ($17.2 < $ log $
N(HI) < 20.3$ atoms cm$^{-2}$) are deduced from the fit to the
observed cumulative number of quasar absorbers (P\'eroux \etal\ 2003),
while the solid bin in the range $19.0 < $ log $ N(HI) < 20.3$ atoms
cm$^{-2}$ is the observed number density of sub-DLAs from the UVES
archive quasar spectra presented in this paper. Models (corrected to
$\Omega_M=0.3$, $\Omega_{\Lambda}=0.7$ cosmology) of Zheng \&
Miralda-Escud\'e (2002) with halo masses of $10^{12}$, $10^{11}$,
$10^{10}$, $10^{9}$ $M_\odot$ (from top to bottom) are
overplotted. The vertical normalisation of the curves is unconstrained
in the models and thus on this figure the lines are arbitrarily
shifted to best fit the data points. On the other hand, the shape of
the model distributions can be directly compared with the data in
order to determine the neutral fraction in the absorber at the radius
where self-shielding starts to be important (see text for further
explanation).}
\label{f:fn}
\end{figure}

Our sample of sub-DLAs from the UVES archive quasar spectra can be
further used to determine the column density distribution of absorbers
in the column density range $10^{19} <$ N(HI) $< 2\times 10^{20}$
atoms cm$^{-2}$. The column density distribution describes the
evolution of quasar absorbers as a function of column density. It is
defined as:

\begin{equation}
f(N, z) dN dX = \frac{n}{\Delta N \sum_{i=1}^{m} \Delta X_i} dN dX
\end{equation}

where $n$ is the number of quasar absorbers observed in a column
density bin $[N, N+\Delta N]$ obtained from the observation of $m$
quasar spectra with total absorption distance coverage $\sum_{i=1}^{m}
\Delta X_i$.  The distance interval, $dX$, is used to correct to
co-moving coordinates and thus depends on the geometry of the
Universe. For a non-zero $\Lambda$-Universe:

\begin{equation}
X(z) = \int_{0}^{z} (1 + z)^2 \left[(1 + z)^2 (1 +
z\Omega_M) - z (2 + z) \Omega_{\Lambda}\right]^{-1/2}dz
\end{equation}
\label{eqn_dist_int}

The values of $dX$ used in our study are given in 
Table 2 for
an $\Omega_M=0.3$, $\Omega_{\Lambda}=0.7$ cosmology. We have sampled a
total redshift path $dz=17.5$. This corresponds to a distance interval
$dX=59.5$ in $\Omega_M=0.3$, $\Omega_{\Lambda}=0.7$ cosmological
model.

We have used the sample of sub-DLAs from the UVES archive quasar spectra to
determine the column density distribution down to N(HI) $= 10^{19}$
atoms cm$^{-2}$ for the first time. The results are shown in
Figure~\ref{f:fn} together with the predictions of P\'eroux \etal\
(2002) computed from the expected number of sub-DLAs and a compilation
of DLAs at all redshifts. The observational results presented here are
consistent with the predicted change of slope of the column density
distribution at N(HI) $\sim$ $10^{19}$ atoms cm$^{-2}$, as suggested
by previous work (Petitjean \etal\ 1993; Storrie-Lombardi \& Wolfe
2000; P\'eroux \etal\ 2003). This feature is probably the signature of
the onset of self-shielding in these systems.

\subsubsection{Comparison with Models}

Although simulating high column density systems such as DLAs is still
extremely challenging because of resolution limitations, Gardner
\etal\ (1997) have tried to overcome the problem by imposing the density
profile of resolved halos onto unresolved ones. They derive the
evolution with redshift of the column density distribution and,
interestingly, they find a flattening of the distribution somewhere in
the region log N (HI) $=18.5 - 20.0$ atoms cm$^{-2}$. Nevertheless, the
theory predicts little change in the form of the column density
distribution function over the redshift range $2-4$. This seems
counter to current observations since a strong evolution of the number
density of quasar absorbers with redshift has been derived (P\'eroux
\etal\ 2002), implying a factor of 3 or more difference in the number
of LLS between $z=2$ and $z=4$.

More importantly, the new set of observations can be directly compared
with models of the column density distribution (Corbelli,
Salpeter \& Bandiera, 2001 and Zheng \& Miralda-Escud\'e, 2002). Zheng
\& Miralda-Escud\'e (2002), in particular, modeled DLAs as spherical
isothermal gaseous halos ionized by external cosmic background to
predict the column density distribution. Their models for different
halo masses: $10^{12}$, $10^{11}$, $10^{10}$, $10^{9}$ $M_\odot$ (from
top to bottom) are overplotted in Figure~\ref{f:fn}. The normalisation
of the curves is arbitrary since it depends, among other things, on
the size distribution of clouds, which is not known, but the shape can
be directly compared with the data. In addition, Zheng \&
Miralda-Escud\'e show that the column density at which the flattening
takes place depends only on the neutral fraction in the absorber at
the radius where self-shielding occurs. The present data exclude the
models with high mass halos.

In addition, according to P\'eroux \etal\ (2002), the change of slope
of the column density distribution should evolve with redshift, moving
out in redshift from $z=2$ to $z=4$.  Unfortunately, the current
sub-DLA sample is not sufficient to study the evolution with redshift
of the distribution down to N(HI) $= 10^{19}$ atoms cm$^{-2}$. More
data are required at high redshift in order to be able to probe the
evolution with time of the properties of the sub-DLAs.

\begin{figure*}
\psfig{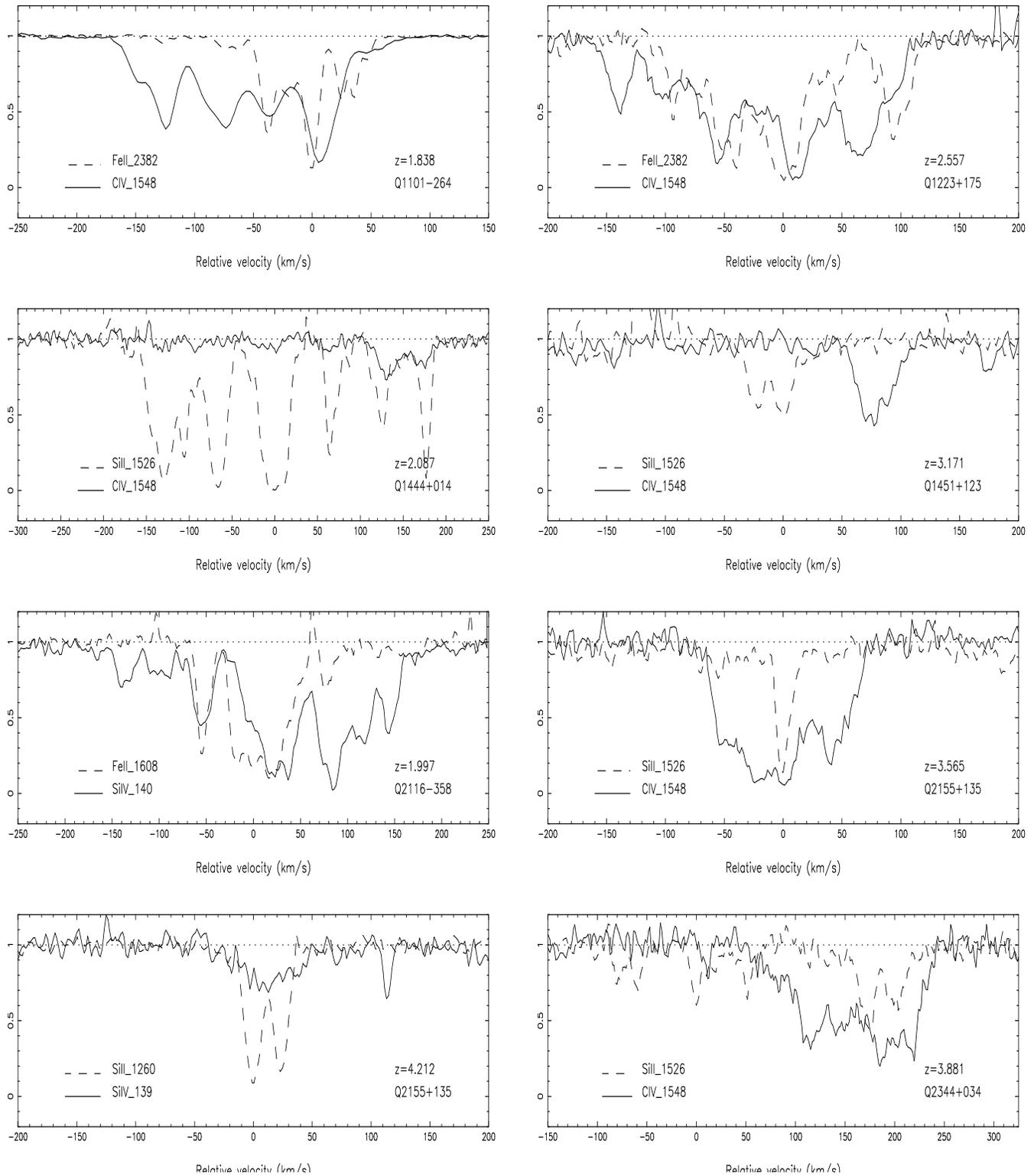}
\caption{Velocity space superposition of the low- (Fe~{\sc ii} or 
Si~{\sc ii}) and high-ionization (C~{\sc iv} or Si~{\sc iv})
transition profiles for the sub-DLAs of our sample where
high-ionization transitions are detected. The two states of the gas
overlap in most cases apart from $z_{\rm abs}=3.171$ in Q1451$+$123. }
\label{f:low_high_shift}
\end{figure*}

\begin{figure*}
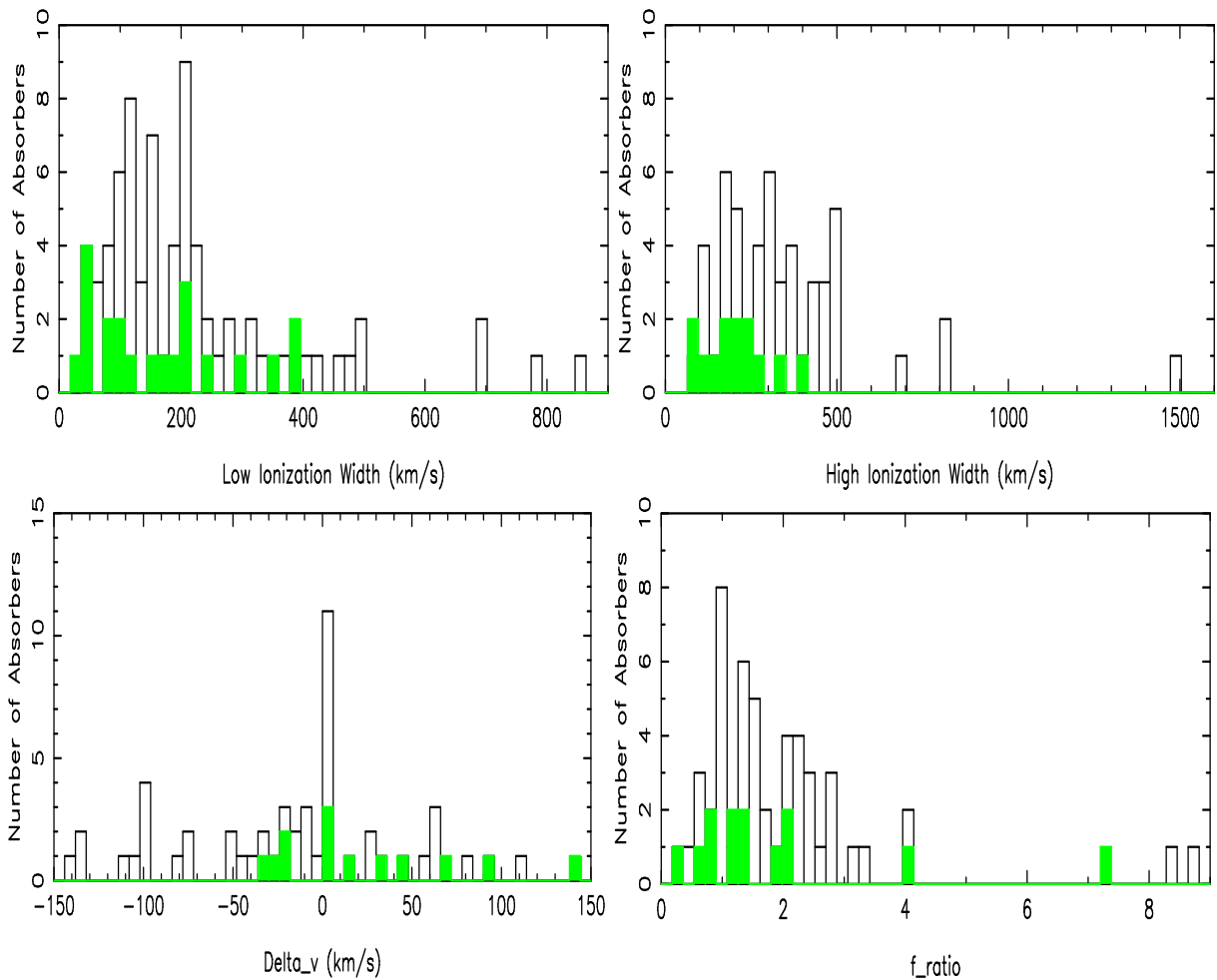

\hspace{0.01cm}
\psfig{figure=peroux_fig3a.ps,angle=-90,width=.45\textwidth,height=.27\textheight}
\psfig{figure=peroux_fig3b.ps,angle=-90,width=.45\textwidth,height=.27\textheight}
\hspace{0.01cm}
\psfig{figure=peroux_fig3c.ps,angle=-90,width=.45\textwidth,height=.27\textheight}
\psfig{figure=peroux_fig3d.ps,angle=-90,width=.45\textwidth,height=.27\textheight}
\caption{Comparison of the kinematic properties of DLAs (black line) 
and sub-DLAs issued from both our sample and the literature (light
coloured histogram). The parameters plotted are (from left to right,
from top to bottom) $\Delta_{\rm low}$, $\Delta_{\rm high}$, $\delta
v$ and $f_{\rm ratio}$. Although the statistics on sub-DLAs are still
small it seems that their properties roughly span the parameter space
of the DLAs, thus challenging multicomponent semi-analytical models.}
\label{f:kine}
\end{figure*}

\section{Clustering of Sub-DLAs}

As already noted all the quasars studied here were extracted from the
ESO UVES archive and so the large majority were observed because there
was a DLA along their line-of-sight. Although the DLAs themselves were
excluded from the redshift path surveyed in the quasar, it is of
interest to note that on at least one occasion, the sub-DLA 
discovered is situated in the wing of a known DLA: $z_{\rm
sub-DLA}=2.557$ and $z_{\rm DLA}=2.466$ towards Q1223$+$1753, giving
$\Delta z = 0.09$, corresponding to $\Delta v \sim $ 8000 km s$^{-1}$
(see Paper I). In addition, we note that out of the 12 sub-DLAs which
make up our sample, at least 3 systems ($z_{\rm sub-DLA}=1.996$
towards Q2116$-$358, $z_{\rm sub-DLA}=3.565$ towards PSS~J2155$+$1358
and $z_{\rm sub-DLA}=3.882$ towards PSS J2344$+$0342) cannot be fitted
with only one absorption system, suggesting the presence of smaller
column density absorbers in their close vicinity (see Paper I for more
details on these systems). Obviously, these features illustrate the
difficulty in determining the HI column density of sub-DLA systems and
differentiating these objects from a conglomerate of smaller column
density objects. In the particular cases of $z_{\rm sub-DLA}=3.356,
3.589$ and $4.214$ towards BR~J0307$-$4945, we cannot estimate the
size and number of absorbers involved. We refer the reader to section
3.1 of Paper I for discussion of these specific cases.

To summarise, our sample of 12 sub-DLAs contain 4 systems associated
with other absorbers, one of which is a DLA, the remaining being
smaller column density systems. From P\'eroux \e\ (2002), the expected
probability of absorbers with N(H) $> 10^{19.0}$ atoms cm$^{-2}$ is
$n(z)\sim 0.4$ at $z \sim 2.5$, if the absorbers where randomly
distributed. Thus there is only a 1 in 50 chance of finding quasar
absorbers separated by $\Delta z = 0.09$ ($\Delta v
\sim$ 8000 km s$^{-1}$). The observed incidence of clustered absorbers
(1 out of 12) is therefore higher than expected. For comparison, out
of the hundreds of DLAs known today, only 3 cases of such clustering
along a given line-of-sight have been reported in the
literature. Prochaska \& Wolfe (1999) report a pair of DLAs towards
Q2359$-$02 at $z=2.1$ and separated by $\Delta v \sim 5700$ km
s$^{-1}$. Ellison \& Lopez (2001) have presented a pair of (strictly
speaking) DLA/sub-DLA at $z =1.8$ separated by $\Delta v \sim 2000$ km
s$^{-1}$ which are characterised by a low $\alpha/{\rm Fe}$ elemental
ratio. Lopez \e\ (2001) have found 3 DLAs at $z =2.6$ in CTQ 247
separated by $\Delta v \sim 6000$ km s$^{-1}$ which show similar
abundance patterns (Lopez \& Ellison 2003).

The quantitative interpretation of this phenomenon is not
straightforward.  Lyman Break selected galaxies (e.g. Steidel \e\
1998) and Ly-$\alpha$ emitting galaxies (Ellison \e\ 2001) have been
shown to be clustered at high redshifts. Similarly, studies of
transverse clustering from quasars pairs or groups show a highly
significant overdensity of strong absorption systems over separation
lengths from $\sim 1$ to 8 h$^{-1}$ Mpc (D'Odorico, Petitjean \&
Cristiani 2002). Our findings are therefore not surprising in that
respect. In all cases, pairs of sub-DLAs/DLAs are potentially of great
interest since, in such cases, one could reasonably assume that the
incident radiation field is similar for both absorbers, thus providing
further information on the otherwise poorly constrained
photoionization spectrum. Such studies would nevertheless require a
good estimate of HI column density of each system (often requiring
other lines down the Lyman series) and that the metals associated with
each absorber are well characterised.

\section{Kinematics of Sub-DLAs}

The kinematic properties of quasar absorbers can be used to provide
further information on their nature. Using various line diagnostics
to parameterise the symmetry, velocity width and the edge-leading
profiles of the metal lines of 17 DLA systems, Prochaska \& Wolfe
(1998) suggest that the absorbers are fully formed, large, rapidly
rotating galactic discs with $v_{\rm circ}$~$\ga$200 $\rm
km~s^{-1}$. More recently, Wolfe \& Prochaska (2000b) have used a
sample of 35 DLAs (Wolfe \& Prochaska 2000a) and compared them with
semi-analytical models of galaxy formation. They note that none of
the models reproduce the overlap between low- and high-ionization
metal profiles observed in the data. In addition, this scenario is in
stark contrast with the currently favoured hierarchical structure
formation models, where present-day galaxies are assembled from
virialised sub-units over a large redshift range ($z\sim 1-5$).
Other hydrodynamic N-body simulations and semi-analytical models
(Haehnelt, Steinmetz \& Rauch 1998; Maller et al.\ 2001; Maller \e\
2002) have shown that the low ion absorption profiles can equally well
be interpreted as a signature from merging proto-galactic clumps in
collapsing dark matter {\it halos} with small virial velocities
($\sim$ 100 $\rm km~s^{-1}$). In addition, Ledoux \e\ (1998) have used
a sample of 26 DLAs and find a correlation between the asymmetry and
$\Delta V$ for $\Delta V \la 150
\rm km~s^{-1}$. They suggest that this correlation is evidence that
rotation velocities may dominate the narrower metal lines.

\subsection{Sub-DLAs Characteristics}

First we examine the number of components necessary to fit the
low-ionization transitions of the sub-DLAs, noting that out of 12
systems, 5 require more than 10 components. We also note that most of
the remaining systems (another 5 absorbers) are well fitted with only
3 or less components.

We chose to describe the properties of the lines of the low-
and high-ionization transitions in our sample of sub-DLAs with the
help of four simple quantities, namely:

\begin{enumerate}
\item{$\Delta_{\rm low}$: the width of the low-ionization transitions as measured from the most saturated Fe~{\sc ii} or Si~{\sc ii} lines. }
\item{$\Delta_{\rm high}$:  the width of the high-ionization transitions as measured from the most saturated C~{\sc iv} or Si~{\sc iv} lines.}
\item{$\delta v$: the offset between the mean velocity of the high- and low-ionization transitions.}
\item{$f_{\rm ratio}$: the ratio of the velocity widths of the high- and low-ionization transitions.}
\end{enumerate}

All these parameters are used to quantify the differences between
various states of the gas observed in absorption. The shift between
the low- and high-ionization transitions for our sample of sub-DLAs
is shown in Figure~\ref{f:low_high_shift} for the 8 systems where
high-ionization transitions are detected. In most cases, although the
two sets of absorption overlap, the profiles are widely different. In
one particular object ($z_{\rm abs}=3.171$ in Q1451$+$123), the
Si~{\sc ii} is not superimposed at all on the C~{\sc iv} doublet in
velocity space, the two being separated by $\sim 75$ km s$^{-1}$. It
might be that the C~{\sc iv} doublet is actually not associated with
the sub-DLA itself (see Paper I for more details). As in DLAs, the
high-ionization transitions spread over larger velocity intervals than
their low-ionization counterparts, but their correspondence suggests
that the systems are within the same potential well (Wolfe \&
Prochaska 2000b). However, in two cases ($z_{\rm abs}=3.882$
in Q2344$+$034 \& $z_{\rm abs}=2.087$ in Q1444$+$014), the
low-ionization transitions traced by Si~{\sc ii} spread over a larger
velocity interval than the high-ionization transitions.

\subsection{Comparison with DLAs}

In order to compare the kinematic properties of sub-DLAs with DLAs, we
have determined the parameters described in the previous section for
72 DLAs taken from the literature. The systems quoted are from a
mixture of UVES/HIRES and lower resolution spectrographs for
high-redshift targets, and from HST at lower redshifts. Thirteen
additional sub-DLAs are found in the literature. We made the relevant
measurements from the quasar spectra whenever we had them, otherwise
we directly used the original publications to quantify the width of
the ion lines. We preferentially used the Fe~{\sc ii} and C~{\sc iv}
lines whenever possible. In addition, if the data came from the UCSD
HIRES/KeckI survey (Prochaska \& Wolfe 1998, 2000 and 2002; Prochaska,
Gawiser \& Wolfe 2001; Prochaska \e\ 2002), we took the velocities
used for the Optical Depth Method given on the survey's web
page~\footnote{http://kingpin.ucsd.edu/$\sim$hiresdla/} kindly
provided to the general community and maintained by Jason
X. Prochaska.

The resulting distributions for $\Delta_{\rm low}$, $\Delta_{\rm
high}$, $\delta v$ and $f_{\rm ratio}$ are illustrated in
Figure~\ref{f:kine}. Although the number of sub-DLAs is still small,
it appears that their velocity spreads in both low- and
high-ionization transitions are statistically equivalent to the DLAs
(top left and top right panel of Figure~\ref{f:kine}). In most cases,
the shift between the low- and high-ionization transitions is small in
DLAs (less than 20 km s$^{-1}$) (bottom left panel of
Figure~\ref{f:kine}). Finally, the $f_{\rm ratio}$ values of DLAs and
sub-DLAs are comparable (bottom right panel of
Figure~\ref{f:kine}). These measurements are important to constrain
semi-analytical models. Indeed, Maller \e\ (2002) find that a large
fraction of absorbers with log N(HI)$>20.0$ cm$^{-2}$ might be
composed of multiple component discs. In contrast, they predict that
the smaller systems (in particular those with $4 \times 10^{19}<$
N(HI)$<10^{20}$ cm$^{-2}$) will mostly be single discs. Therefore,
$\delta v$ is expected to be smaller in these systems than in
classical DLAs. Although our data set is still small, the observations
presented here do not support this interpretation, thus challenging
multiple component models. A possible explanation is that some of the
low ionization gas in sub-DLAs is in hot gas. Indeed, the data show a
hint of decreasing $f_{ratio}$ with lower column densities, therefore
suggesting that the high and low ions are in the same gas at lower
N(HI).

\begin{figure*}
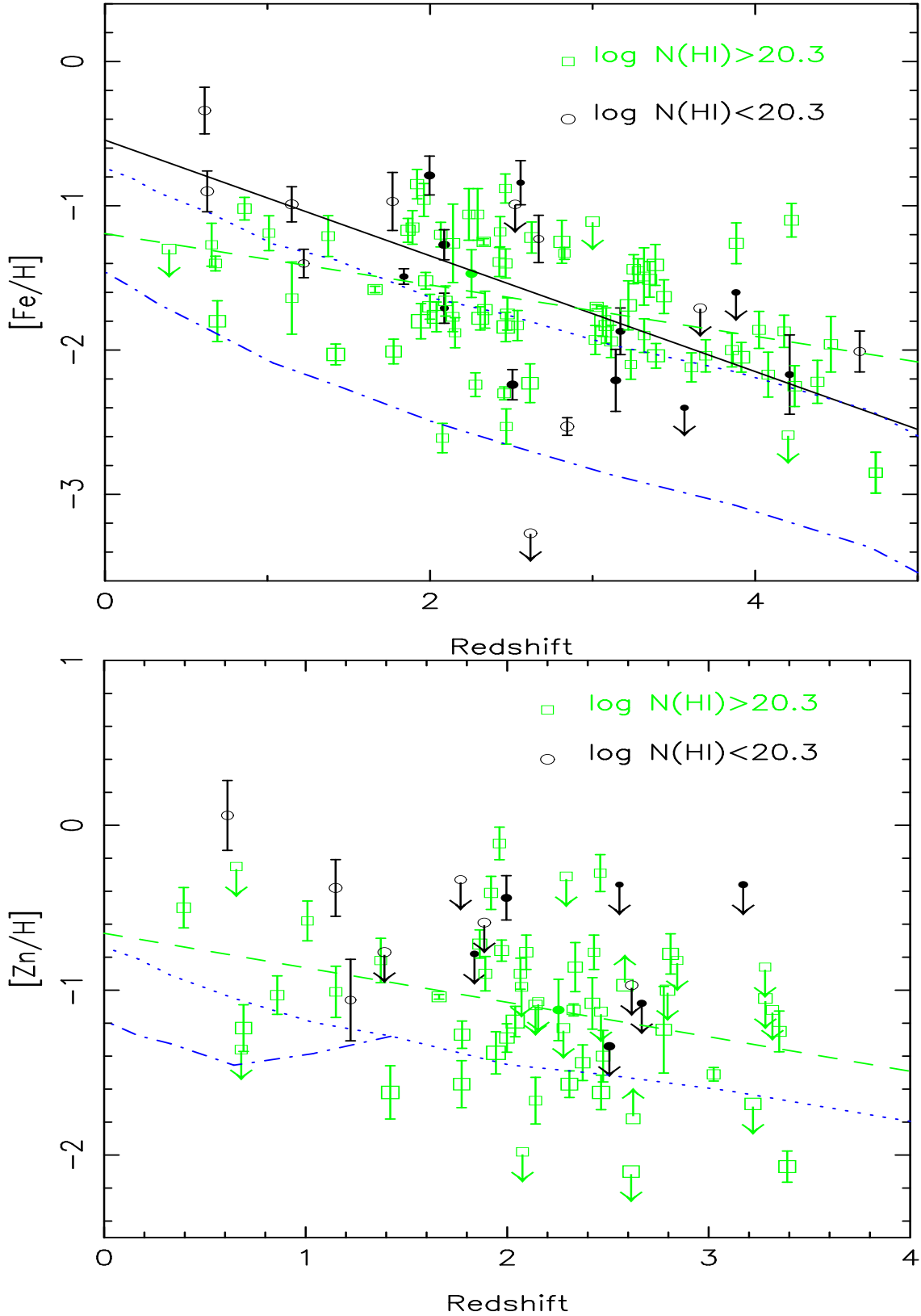

\begin{center}
\psfig{figure=peroux_fig4a.ps,angle=-90,width=.85\textwidth,height=0.45\textheight}
\psfig{figure=peroux_fig4b.ps,angle=-90,width=.85\textwidth,height=0.45\textheight}
\caption{[Fe-Zn/H] metallicities for DLAs (squares) and sub-DLAs (open circles 
for data from the literature, filled circles for the sub-DLAs from
this work) as a function of redshift. The symbol sizes are
proportional to the HI column density of the systems. The evolution
with redshift of the [Fe/H] ratio is more pronounced for sub-DLAs than
for DLAs (solid and dashed lines, respectively). The dashed-dotted
(dotted) curves are results from models from Hou, Boissier \& Prantzos
(2001) for sub-DLAs (DLAs).}
\label{f:Fe_z}
\end{center}
\end{figure*}

\section{Chemical Abundances of Sub-DLAs}

\subsection{Ionization Fraction}

In Paper I, we have presented the hydrogen and ionic column densities
of all of the sub-DLAs included in Table~\ref{t:sub-DLA_stat}, using
standard Voigt profile fitting. Since their relatively low column
density implies that some of the gas might be ionized, the ionization
state of the sub-DLAs has been investigated in detail in Paper I
using the photoionization models of the CLOUDY software package. The
determination of the ionization parameter $U$ for a given absorber
system crucially depends on the detection of intermediate-ionization
transitions. In addition, the photoionization model relies on several
input parameters related to the nature of the absorber and the
characteristics of the radiation field. In this context we chose to
extract from our analysis a global ionization correction rather than
individual corrections limited by partial information. We found the
correction to be $<0.3$ dex for most elements down to N(HI)$= 19.3$
cm$^{-2}$. We also note that the correction to apply is quite
different from species to another. In the case of O~{\sc i} or C~{\sc
ii}, for example, the ionization fraction is small, while it becomes
important for such elements as Al~{\sc ii} or Zn~{\sc ii}. One needs to
keep these facts in mind while interpreting the various figures
presented in the following sections.

\subsection{Metallicity Evolution}

In this section, we present the results of the metallicity study of
the sample of sub-DLAs and compare them with the characteristics of
the well studied DLAs. The references for absorbers from the
literature are the same as in Section 4.2. The chemical abundances in
DLAs have been used to trace the metallicity evolution of galaxies
over a large look-back time (e.g. Pettini et al. 1999; Prochaska \&
Wolfe 2000) and contrary to virtually all chemical models, the
observations in DLAs indicate mild evolution with redshift. By
extending the abundance determination to systems of lower column
densities, such as the sub-DLAs, we can gain insight in the evolution
of the galactic chemical evolution of galaxies, since they are more
likely to represent the basic building blocks of hierarchical growth
of structure.

\subsubsection{Metallicity of Individual Systems}

\begin{table}
\begin{center}
\caption{Results from the order ranking Kendall test and slopes of the 
fits assuming a linear correlation for different sub-sets of quasar
absorbers, ignoring all upper and lower limits. The lack of
correlation between the redshift and [Fe/H] is due to the larger
column density systems (log N(HI) $>$ 21.0). In contrast the
sub-DLAs appear to show more correlation, with a more pronounced
slope, although the results are still weakened by low number
statistics.}
\label{t:kendall}
\begin{tabular}{lcccc}
\hline
data set &nber of&$\tau$ &c.l. &Slope\\ 
definition & abs&&&\\
\hline
log N(HI) $\geq$ 20.3	&72	&--0.29		&99.53\%	&$-0.18\pm0.12$\\
log N(HI) $<$ 20.3	&17	&--0.50		&84.27\%	&$-0.40\pm0.22$\\
log N(HI) $<$ 21.0	&75	&--0.40		&99.99\%	&$-0.28\pm0.11$\\	
\hline
All Absorbers		&89	&--0.35		&99.53\%	&$-0.24\pm0.11$\\
\hline
\end{tabular}
\end{center}
\label{t:nz}
\end{table}

{\bf [Fe/H]:} Given that the ionization correction on the Fe~{\sc
ii} column density is within the observational errors, the abundance
measurements of the sub-DLAs can be directly compared with those of
the DLAs gathered from various sources in the literature. We use the
abundance ratios with respect to solar values defined in the usual
manner~\footnote{$[X/H] = \log [N(X)/N(H)]_{DLA}- \log
[N(X)/N(H)]_{\sun}$, assuming that $N(X) =N (X II)$ and $N (H) = N (H
I)$.} and presented in Paper I. For the sake of homogeneity we
recompute these ratios for all data from the literature assuming
[Fe/H]$_{\sun}=-4.50$ (Grevesse \& Sauval 1998). The top panel of
Figure~\ref{f:Fe_z} presents the evolution with redshift of the [Fe/H]
absolute abundances in individual systems. The symbol sizes in this
figure are proportional to the HI column density of the systems. The
classical DLAs are represented with square symbols while the sub-DLAs
are circles (open for data from the literature, filled for the data
from the sample presented in this paper). We applied the order ranking
Kendall test to several subsets of the sample as shown in
Table~\ref{t:kendall}, ignoring all upper and lower limits. As
previously reported (e.g. Prochaska \& Wolfe 2002; Dessauges-Zavadsky
\e\ 2001), there is evidence for a mild evolution of the individual [Fe/H]
in DLAs (correlation coefficient of --0.29 at 99.53\% confidence
level). The analysis suggests a stronger correlation of the [Fe/H] in
sub-DLAs although the significance of the test is weakened by low number
statistics (correlation coefficient of --0.50 at 84.27\% confidence
level for a total of 17 sub-DLAs). The effect is predominant at $z<2$
and most of the evolution observed lies in this redshift range. We
fitted the two populations of absorbers assuming a linear correlation
and found a slope of $\alpha=-0.18\pm0.12$ for DLAs and
$\alpha=-0.40\pm0.22$ for sub-DLAs (dashed and solid line of the top
panel of Figure~\ref{f:Fe_z} respectively). The evolution with
redshift of the [Fe/H] ratio might be more pronounced for sub-DLAs
than for DLAs. This difference, if confirmed by a larger sample of
data, suggests that the detection of DLAs is more biased by dust at
low redshift, or that the sub-DLAs are associated with a class of
galaxies which better trace the overall chemical evolution of the
Universe.

{\bf [Zn/H]:} The well-known drawback of using Fe~{\sc ii} for
abundance determination is the fact that this element is sensitive to
depletion onto dust grains (Pettini \e\ 1997). An alternative to
overcome the dust depletion issue is provided by Zn, an element known
to be only mildly depleted. The lower panel of Figure~\ref{f:Fe_z}
present the evolution of the [Zn/H] with redshift. Again for the sake
of homogeneity we recompute these ratios for all data from the
literature assuming [Zn/H]$_{\sun}=-7.33$ (Grevesse \& Sauval
1998). In our sample of sub-DLAs, most of the Zn~{\sc ii} lines are
weak or undetected and thus we can only provide upper limits. In
addition, the ionization correction in sub-DLAs is important for
Zn~{\sc ii} and thus the interpretation of the abundances is more
complicated. We again used the order ranking Kendall test for the
subsets of the Zn~{\sc ii} sample. When considering both DLAs and
sub-DLAs (and excluding all limits), we determine a slope
$\alpha=-0.28\pm0.23$, i.e. slightly shallower than the one derived by
Vladilo \e\ (2000) and Dessauges-Zavadsky \e\ (2001) ($\alpha \sim
-0.3$) since the latter works loosely include sub-DLAs in their
samples. Nevertheless, the significance of this result is low
(correlation coefficient of --0.25 at 84.27\% confidence level for a
total of 40 abundances). Analysing DLAs only, leads to a slope
$\alpha=-0.21\pm0.25$ for 36 systems. There are not enough Zn~{\sc ii}
measurements in the sub-DLAs range to allow such analysis. An
important point, however, is the fact that the slope in the Zn~{\sc
ii} DLA sub-set ($\alpha=-0.21\pm0.25$), a tracer of true metallicity
evolution, is not as steep as in the Fe~{\sc ii} sub-DLA sub-set
($\alpha=-0.40\pm0.22$). This suggests that the evolution we probe in
these latter absorbers is {\it not} due to the hidden effect of dust.

\begin{figure*}
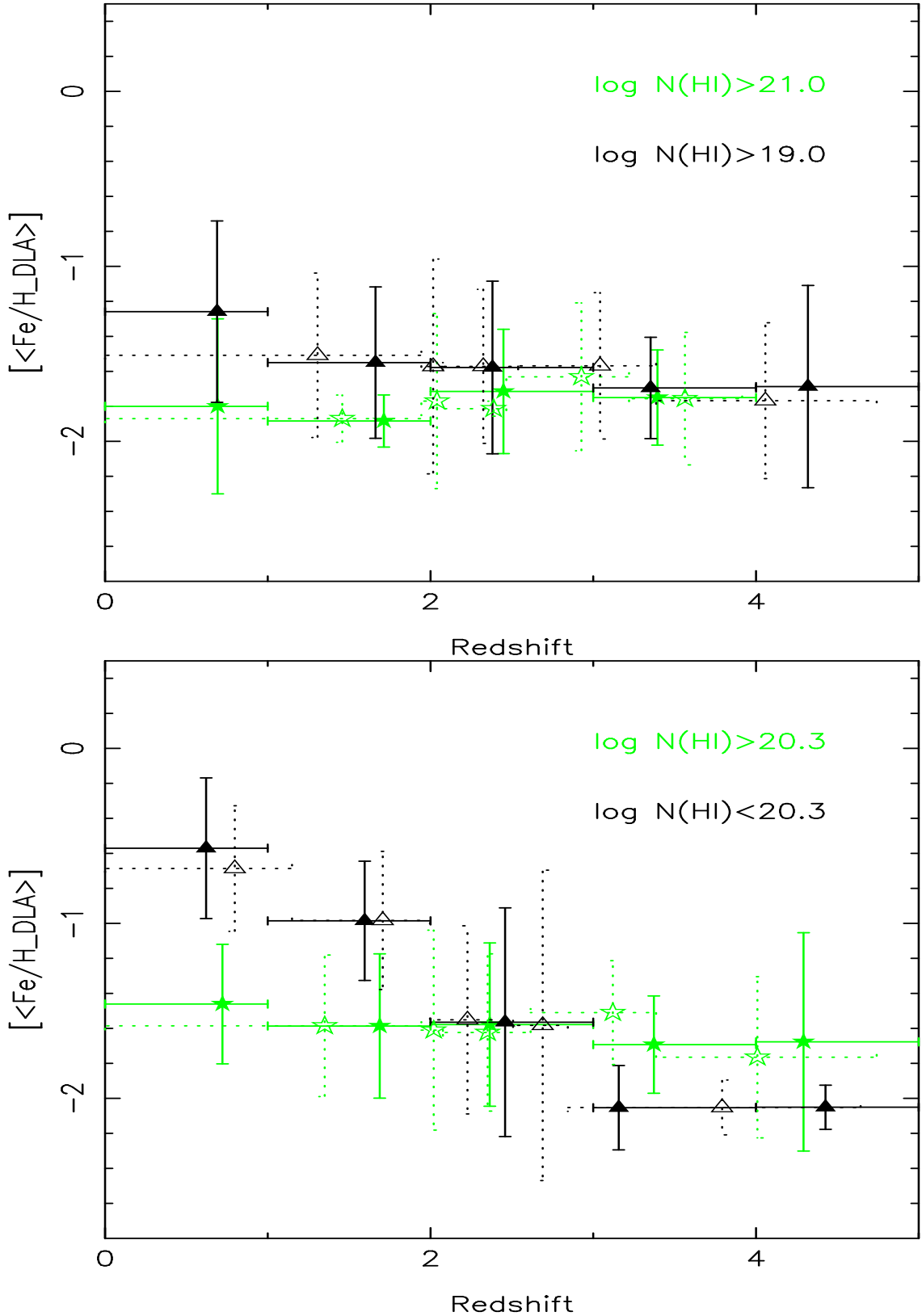

\psfig{figure=peroux_fig5a.ps,angle=-90,width=.85\textwidth,height=0.45\textheight}
\psfig{figure=peroux_fig5b.ps,angle=-90,width=.85\textwidth,height=0.45\textheight}
\caption{HI column density-weighted mean metallicities for various sub-sets of quasar absorbers. The dotted bins are for constant HI intervals
and the solid bins are for constant redshift intervals. The evolution
of $[\langle \rm Fe/\rm H_{\rm DLA}\rangle]$ is clearly more
pronounced for sub-DLAs than for DLAs, but this result is not apparent
when all absorbers with N(HI) $>10^{19.0}$ atoms cm$^{-2}$ are
considered. Indeed, in our sample the number of DLAs (72) is much
larger than the number of sub-DLAs (17). 
}
\label{f:weigth_mean}
\end{figure*}

{\bf Comparison with Models:} These observational results can be
compared with models of galaxy evolution and the corresponding
evolution of metal abundances to further test the nature of quasar
absorbers. Complications arise from the fact that the observed systems
might have different formation epochs, a parameter which cannot be
directly constrained by the data. Prantzos \& Boissier (2000) use
models of the chemical evolution of disc galaxies, calibrated on the
Milky Way (Boissier \& Prantzos 1999, 2000), which they then compare
with the Zn abundances of DLAs. This comparison is extended to other
elements in Hou, Boissier and Prantzos (2001). They find that the
results are compatible with available observations, provided that
several biases are taken into account, one of these biases being
reproduced by applying a ``dust filter'' to the models. This filter
was proposed by Boiss\'e et al. (1998) who suggest that high column
density and high-metallicity systems are not detected because the
light of background quasars is severely extinguished. This claim
arises from the observed trend in [X/H] versus log N(HI) in quasar
absorbers (see section 5.2.3 for more on this issue). The results of
the models are further constrained by the HI column density detection
limits in current surveys. In Hou, Boissier \& Prantzos (2001), this
limit corresponds to the formal definition of DLAs, $\log \rm N(HI) >
20.3$. The mean [Fe/H] value of the model is presented in
Figure~\ref{f:Fe_z} as a dotted curve. The scatter in the observations
is genuine (since it is higher than the error bars on individual
metallicity measurements) and is well reproduced by the model (see
Figure~2 of Hou, Boissier \& Prantzos 2001).

Adjusting the threshold to the sub-DLA definition results in the
dashed-dotted curve presented in Figure~\ref{f:Fe_z}. The models do
not indicate a change in evolution with redshift between DLA and
sub-DLA systems (see dotted and dashed-dotted curves in the bottom
panel of Figure~\ref{f:Fe_z}). Rather they suggest that the
metallicity in sub-DLAs is lower than in DLAs. This is a natural
consequence of the abundance gradients observed in nearby galaxies and
in the Milky Way, well reproduced by the models (Prantzos \& Boissier
2000). On the contrary, the observations indicate that the sub-DLAs
seem more metal-rich than the DLAs, at least at low
redshift. Therefore although the disc models are in quite good
accordance with the DLAs observations, the mismatch between these
models and the sub-DLA observations suggests that sub-DLAs do not
arise in disc galaxies.

The [Zn/H] observed redshift evolution can also be compared with
models of galaxy evolution proposed by Hou, Boissier \& Prantzos
(2001). It should be emphasized that in the models, the Zn~{\sc ii}
abundance is set to follow the Fe abundance. The difference between
the two arises only from different ``biases''; the current Zn
detection can be accounted for by applying to the models the filter:
log N(HI)+[Zn/H]$>$18.8, an estimate of the Zn limit detection
(e.g. Prantzos \& Boissier 2000). Again, setting the column density
cut-off in the models to the DLA definition reproduces the trend well
(dotted curve in the bottom panel of Figure~\ref{f:Fe_z}) and
the scatter between measurements (see also Prantzos \& Boissier
2000). Once the cut-off is chosen to match the sub-DLA definition
(dashed-dotted curve in bottom panel of Figure~\ref{f:Fe_z}), the
models suggest that the Zn~{\sc ii} abundances in the systems at
redshift greater than $\sim$ 1.3 are below current detection
limits. This is indeed what is observed since most of the [Zn/H]
presented here are upper limits. Note that to be able to measure the
Zn abundance (log N(HI)+[Zn/H]$>$18.8) in sub-DLAs, the system must
satisfy : [Zn/H] $>$ -1.5.  

An alternative modelling approach has been recently proposed by Cen
\e\ (2003) using hydrodynamic simulations to reproduce the gas and
metal evolution observed in DLAs. They find that the slow metallicity
evolution currently observed in DLAs can be explained by the sequential 
formation of galaxies from the highest metallicity systems. This
effect would be combined with the increase of metallicity of the
absorbers with the smallest abundances. Including
an artificial cut-off in order to mimic the presence of dust in DLAs,
the simulations are found to be in excellent agreement with the
observations (Cen \e\ 2003).

\subsubsection{Weighted Mean Metallicity}

Since DLAs and sub-DLAs are known to contain a major fraction of the
neutral gas at all redshifts (P\'eroux \e\ 2003), all of these
systems should be used to probe the metallicity of the Universe and
its evolution with time. A quantitative way of estimating this
evolution is provided by the HI-weighted mean metallicity. Pettini \e\
(1997) state that ``under the working assumption that DLAs account for
most of the material available for star formation at high redshift,
the quantity $[\langle \rm X/\rm H_{\rm DLA}\rangle]$ (X=Fe or Zn) is
a measure of the degree of metal enrichment by the Universe at a given
epoch''.

Figure~\ref{f:weigth_mean} presents the column density-weighted mean
Fe~{\sc ii} abundances of $n$ systems in redshift bins of $\Delta z=1$
(solid bins) and in constant HI interval (dashed bins), for various
sub-samples of quasar absorbers. Although more advanced statistical
methods have been recently suggested (Kulkarni \& Fall 2002), we
choose to follow the prescription from Pettini \e\ (1997):

\begin{equation}
	{\rm [} \langle {\rm Fe/H}_{\rm DLA}\rangle {\rm ]} = 
        {\rm log}\langle{\rm(Fe/H)}_{\rm DLA}\rangle-{\rm log~(Fe/H)}_{\odot} 
	\label{}
\end{equation}
where
\begin{equation}
	\langle{\rm (Fe/H)}_{\rm DLA}\rangle = 
        \frac{\sum\limits_{i=1}^{n} N{\rm(FeII}{\rm)}_i}
        {\sum\limits_{i=1}^{n} N{\rm(H_{total}}{\rm)}_i}
	\label{}
\end{equation}

and the error are estimated from the standard deviation,
$\sigma^{\prime}$:

\begin{equation}
	 \sigma^{\prime 2} = 
        \left(\sum\limits_{i=1}^{n}
        \lgroup {\rm[Fe/H]}_i - 
         {\rm [} \langle{\rm Fe/H}_{\rm DLA}\rangle {\rm ]}\rgroup ^2\right)/(n - 1)
	\label{}
\end{equation}
with
\begin{equation}
	{\rm [} {\rm Fe/H}_i {\rm ]} = 
        {\rm log}{\rm(Fe/H)}_i-{\rm log~(Fe/H)}_{\odot} 
	\label{}
\end{equation}

The results of these calculations are shown in
Figure~\ref{f:weigth_mean} for different subsets of quasar
absorbers. The top panel includes all quasar absorbers (DLAs +
sub-DLAs) as well as systems with N(HI) $>10^{21.0}$ atoms
cm$^{-2}$. The bottom panel shows DLA and sub-DLAs separately. We note
that the evolution of $[\langle
\rm Fe/H_{\rm DLA}\rangle]$ is clearly more pronounced for sub-DLAs than for
DLAs, but this result is not apparent when all absorbers with N(HI)
$>10^{19.0}$ atoms cm$^{-2}$ are considered. Indeed, in our sample the
number of DLAs (72) is much larger than the number of sub-DLAs
(17). On the contrary, in the Universe we expect the number of
sub-DLAs to be up to 3 times (at $z\sim 4$) the number of DLAs
(P\'eroux \e\ 2002). Since high column density systems dominate the
HI-weighted mean metallicity, it will be important to increase the
sub-DLA sample size to better probe $[\langle \rm Fe/\rm H_{\rm
DLA}\rangle]$. Nevertheless, these results show that the HI column
density-weighted mean metallicity Fe~{\sc ii} of sub-DLAs {\it do}
evolve with redshift more markedly than for the DLA population. The
dotted bins correspond to constant HI intervals and therefore present
similar number of systems. They also reveal an increase of
metallicity with decreasing redshift, showing that this result is not
an artifact of the low number of sub-DLAs known at $z<2$. Starting
from a metallicity 1/100 solar at $z \sim 4.5$, the sub-DLAs evolve up
to a metallicity 1/3 solar at $z \sim 0.5$. The ionization correction
cannot explain the evolution observed since we have shown in Paper I
that for [Fe~{\sc ii}/HI], it {\it does not} exceed 0.2 dex. These
results again reinforce the hypothesis that sub-DLAs better trace the
global metallicity evolution and therefore should be included in the
metallicity determination to obtain a complete
picture of the abundance.

\begin{figure*}
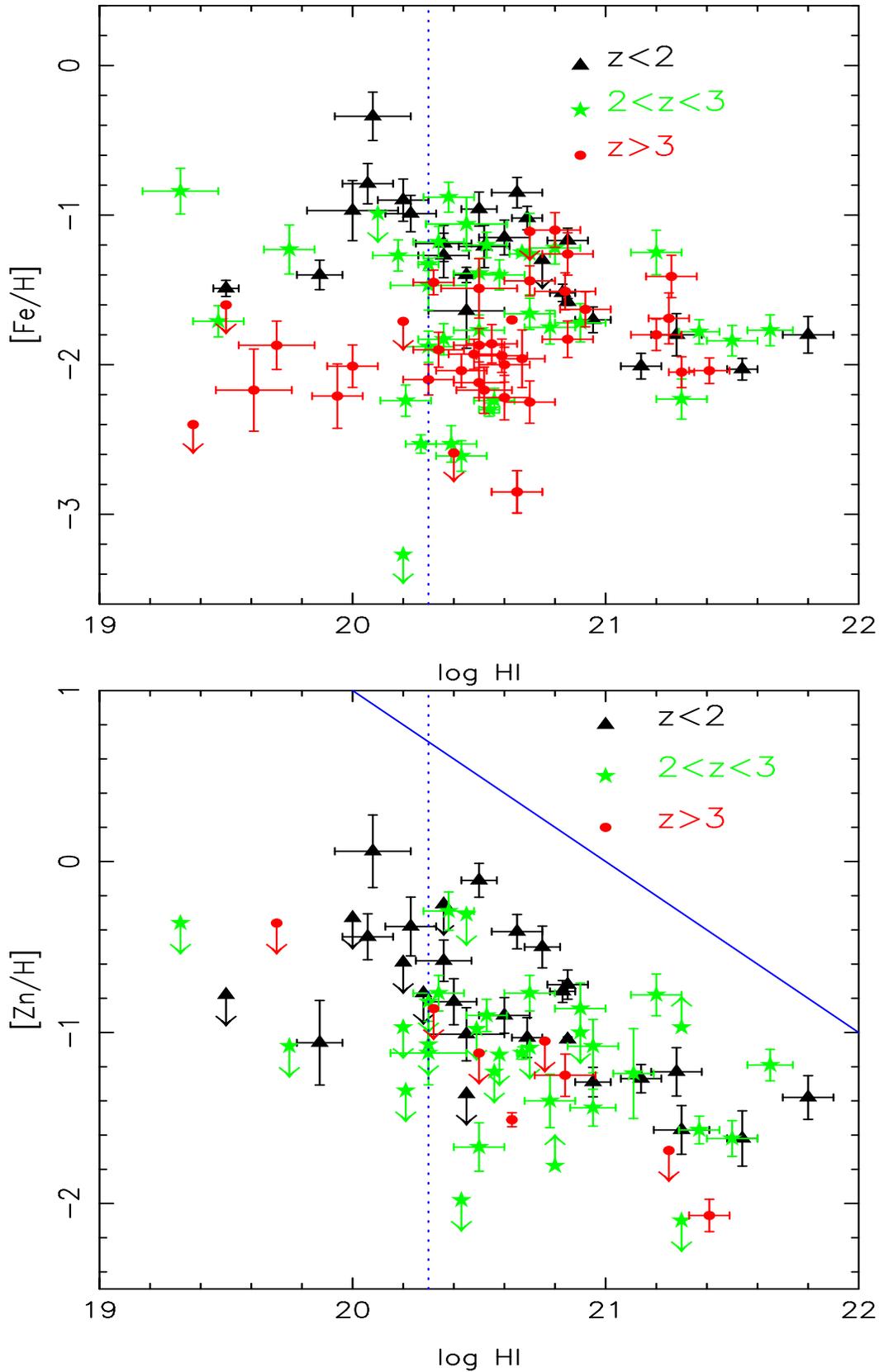

\psfig{figure=peroux_fig6a.ps,angle=-90,width=140mm,height=110mm}
\psfig{figure=peroux_fig6b.ps,angle=-90,width=140mm,height=110mm}
\caption{[Fe/H] and [Zn/H] as a function of HI column density in 
DLAs and sub-DLAs. The dotted line marks the DLA formal definition. On
the bottom panel, the solid line corresponds to the ``dust filter''
proposed by Boiss\'e et al. (1998), namely [Zn/H]+logN(HI) $>$21.}
\label{f:Fe_HI}
\end{figure*}

\begin{figure*}
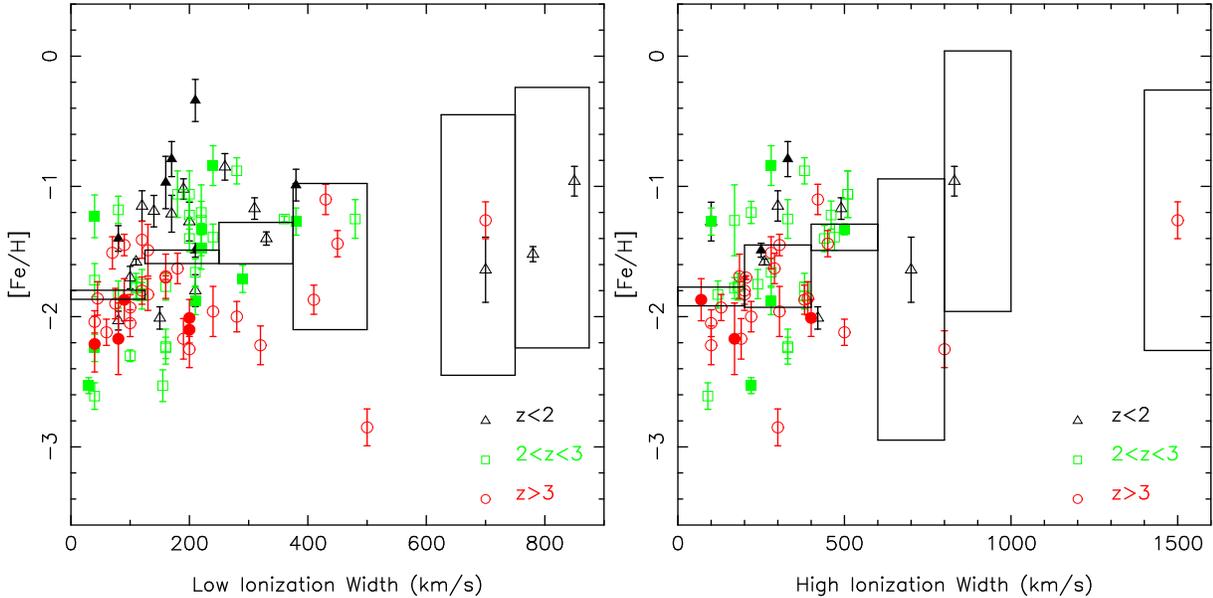

\hspace{0.01cm}
\psfig{figure=peroux_fig7a.ps,angle=-90,width=.45\textwidth,height=.33\textheight}
\psfig{figure=peroux_fig7b.ps,angle=-90,width=.45\textwidth,height=.33\textheight}
\hspace{0.17cm}	  
\caption{[Fe/H] as a function of various kinematic parameters: 
$\Delta_{\rm low}$ and $\Delta_{\rm high}$ (see Section 3 for
details). The symbol shapes depict different redshift ranges. The
open symbols are for DLAs and the filled symbols are for sub-DLAs. The
boxes represent the mean in a given velocity interval with {\it rms} errors
and suggest an increase of metallicity towards larger ionization
widths.}
\label{f:Fe_delta}
\end{figure*}

\subsubsection{Correlating with Other Properties}

{\bf [Fe-Zn/H] versus HI:} Figure~\ref{f:Fe_HI} presents [Fe-Zn/H] as
a function of HI column density for DLAs and sub-DLAs in three
different redshift ranges: $z<2$, $2<z<3$ and $z>3$. In the top panel
([Fe/H]), the spread of values is larger at lower HI column
densities. This suggests that we are sampling a different class of
objects, or at least different parts of the same object as suggested by
e.g. Boissier, P\'eroux \& Pettini (2002). Another possibility is
that dust plays an important role in the diversity of properties through
a range of star
formation histories. In contrast, there is little spread in the
[Fe/H] ratio at very high HI column densities (log N(HI)$>21$) posing the
question of why these systems differ from the
other quasar absorbers. This might indicate that systems with high
column densities belong to a well-defined population of objects, though
more probably,
the lack of apparent spread could be the result of a bias against high
column density, high metallicity absorbers due to their high dust
content (Boiss\'e \e\ 1998). Indeed, the bottom panel of
Figure~\ref{f:Fe_HI} shows that the observational bias noted by
Boiss\'e et al. in the HI versus [Zn/H] relation, still holds (i.e. no
data are situated above the [Zn/H]+logN(HI) $>$21 solid line). A
direct dust diagnostic in sub-DLAs is not straightforward to make
since Zn~{\sc ii} is seldom detected. Using the three Zn~{\sc ii} detections
in our sample of 12 systems, we find that the [Zn/Fe] value
is around 0.3-0.4 dex. This suggests that the amount of dust is not
negligible, although naturally there are too few detections to draw any
firm conclusions. In addition, a study of absorbers derived from a
radio-selected quasar sample (CORALS) does {\it not} indicate that a
dust bias greatly affects DLAs studies (Ellison \e\ 2001), but in the
authors own words the conclusions are only tentative due to the
limited size of the data set. Furthermore, in the range $20.0<$log
N(HI)$<21.0$, i.e. the bulk of the data available, the high redshift
objects are less evolved than the others. In any case, we find that
the metallicity evolution of quasar absorbers is a strong function of
the HI column density. In addition, we have shown that the ionization
correction in Fe~{\sc ii} (see Paper I) is similar within a group of
systems with similar column densities. In the sub-DLA range, it
tends to increase the Fe~{\sc ii} column density measured by 0.1-0.2,
thus not affecting the apparent evolution.

{\bf [Fe-Zn/H] versus kinematic properties:} Figure~\ref{f:Fe_delta}
presents the metallicity of quasar absorbers as a function of their
kinematic properties ($\Delta_{\rm low}$, $\Delta_{\rm high}$) as
measured in Section 3. The symbol shapes denote different redshift
ranges. The open symbols are for DLAs and the filled symbols are for
sub-DLAs. The results indicate that in most cases the properties are
homogeneously mixed between samples of different HI column densities
or redshifts. The boxes represent the mean in a given velocity
interval with rms errors and suggest an increase of metallicity
towards larger ionization widths, although the statistical
significance of this result is low.

\subsection{Relative Abundances}

\begin{figure*}
\centering
\psfig{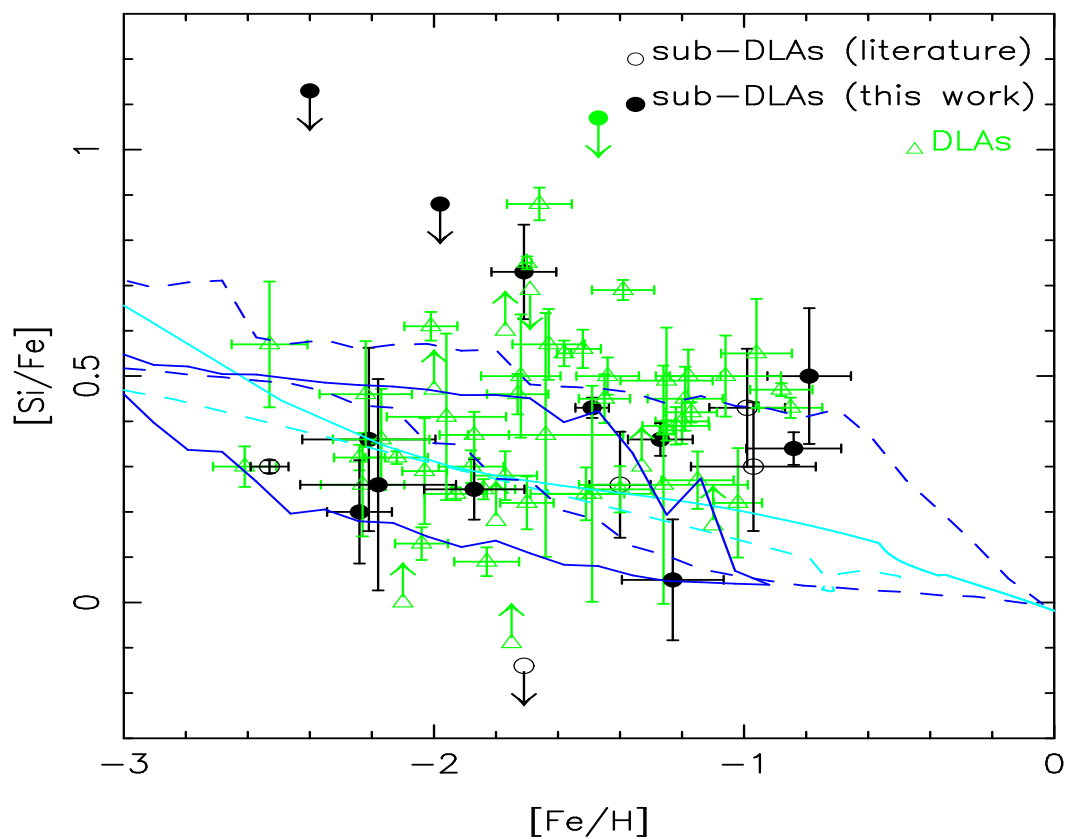}
\caption{[Si/Fe] ratio as a function of metallicity. [Si/Fe] is similar 
in both DLAs and sub-DLAs. The dark-coloured solid (dashed) curves are
contours for sub-DLAs (DLAs) column densities derived from models of
disc galaxies (Hou, Boissier \& Prantzos 2001). The light-coloured
solid (dashed) curves are for spirals at 20 kpc 
(irregulars)
representative of sub-DLAs (DLAs) in the models of (Calura, Matteucci
\& Vladilo 2003). }
\label{f:SiFe}
\end{figure*}

\begin{figure*}
\psfig{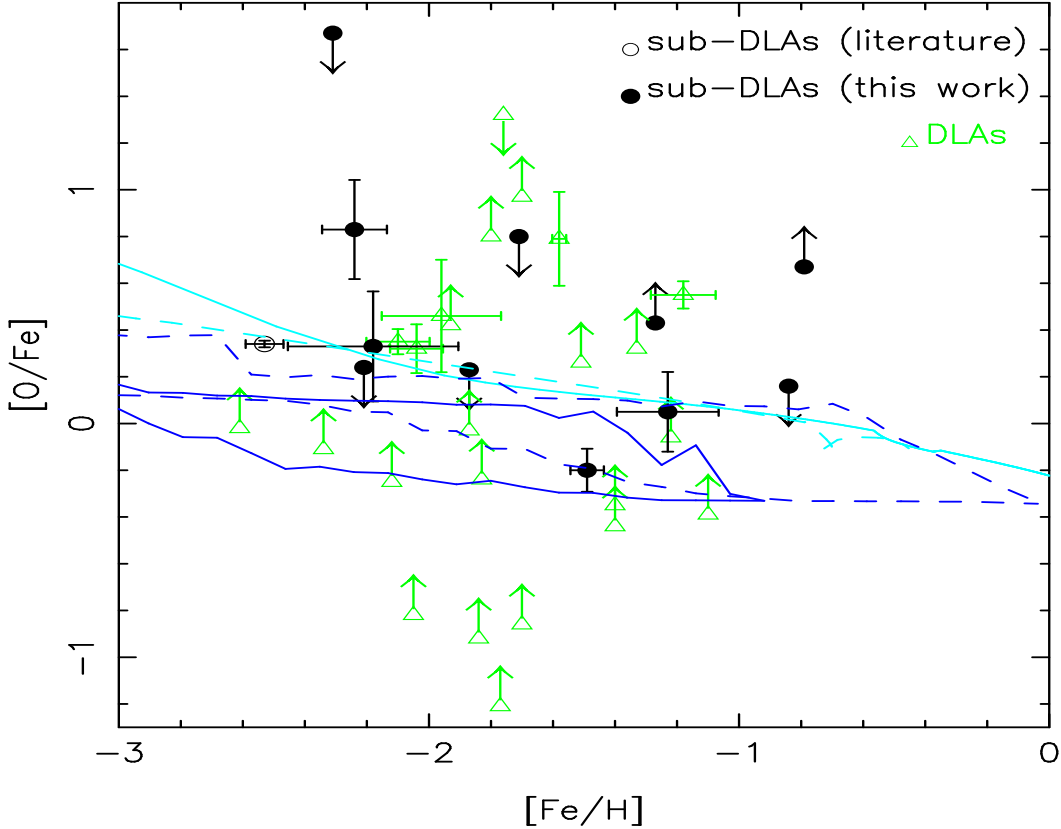}
\caption{[O/Fe] ratio as a function of metallicity.  At low metallicity 
the ratio are greater than solar. Sub-DLAs generally have low
[O/Fe], some of which are sub-solar. The dark-coloured solid (dashed)
curves are contours for sub-DLAs (DLAs) column densities derived from
models of disc galaxies (Hou, Boissier \& Prantzos 2001). The
light-coloured solid (dashed) curves are for spirals (irregulars)
representative of sub-DLAs (DLAs) in the models of (Calura, Matteucci
\& Vladilo 2003). }
\label{f:OFe}
\end{figure*}

The comparison of elemental ratios, independent of HI column density,
but as a function of metallicity is also very important. We concentrate in
particular on the $\alpha$ (i.e. O, Si) over Fe-peak element ratios
versus [Fe/H] metallicities, since this relation provides information
on the star formation histories of quasar absorbers. [$\alpha$/Fe]
versus [Fe/H] is a strong function of the star formation history:
$\alpha$/Fe depends on the lifetimes of the element progenitors,
whereas [Fe/H] depends on the star formation rates. The
$\alpha$-elements are typical products of Type II supernovae, which
are likely to dominate the early stages of galaxy formation. Fe-peak
elements on the contrary are products of Type Ia supernovae, which
play a role in later evolutionary stages. Therefore, high overabundances
of $\alpha$-elements at low [Fe/H] indicate recent star formation,
while solar $\alpha$-element abundances at low [Fe/H] suggests a
lower star formation rate or bursts separated by quiescent
periods.  Nevertheless, the interpretation of these elemental ratios is
once again complicated by differential depletion: Fe~{\sc ii} and
Si~{\sc ii} are known to be depleted onto dust grains in different
proportions while O~{\sc i} is expected to remain unaffected by dust
depletion.

{\bf [Si/Fe]:} Figure~\ref{f:SiFe} shows that [Si/Fe] is similar in
both DLAs and sub-DLAs, averaging around +0.5 dex. For the sake of
homogeneity we recompute these ratios for all data from the literature
and models assuming [Si/H]$_{\sun}=-4.44$ (Grevesse \& Sauval
1998). This value is typical of Galactic metal-poor stars and is a
strong indicator of $\alpha$ enhancement if dust depletion does not
play a major role. Nevertheless, the differential depletion of these
two elements can lead to the same enhancement as the one
observed. This observation can be directly compared with various
models of galaxy formation and evolution. Models of Hou, Boissier \&
Prantzos (2001), with cut-off at the DLA and sub-DLA column density
definitions, are presented in Figure~\ref{f:SiFe} as dark-coloured
dashed and solid contours, respectively. In both cases, the models
seem to underproduce the Si~{\sc ii } abundance with respect to Fe in
comparison with observations, but since the yields used in the model
are uncertain by a factor 2, at best, it is unlikely that the absolute
value of the abundance ratio is known with an uncertainty better than
0.6 dex.

In an alternative approach, Calura, Matteucci \& Vladilo (2002) have
produced models for a variety of galaxy types. In Figure~\ref{f:SiFe},
the light-coloured solid (dashed) curves are for spirals
(irregulars). Indeed, in the chemical evolution models, irregulars are
found to reproduce well the observed properties of DLAs. In these
models, there is no artificial bias applied to mimic the effect of
dust on the observations, nor was it necessary to add an extra HI
column density cut-off to match the quasar absorbers' definition. In
the case of spiral galaxies, in particular, the models presented in
Calura, Matteucci \& Vladilo (2002) never reach the sub-DLA column
density since star formation is set to occur only above a density
threshold, as suggested by Chiappini \e\ (1997). In order to reproduce
the properties of sub-DLAs, it has been necessary to eliminate this 
threshold and extend
the computations towards larger radii (20 kpc). For [Si/Fe], the two
types of models show little difference. In addition, the
interpretation of this elemental ratio might be complicated by the
effect of dust.

{\bf [O/Fe]:} O~{\sc i} is not often measured in DLAs, since the most
predominant line, O~{\sc i} 1302, is often saturated. In contrast,
in sub-DLAs where the column densities are considerably smaller, this
line is well suited for the abundance determination of oxygen. This
element, which is not affected by dust depletion, is also a typical
product of Type II supernovae. The prospect of measuring O~{\sc i} in
large samples of sub-DLAs means that we will have a good indicator of
$\alpha$ over Fe-peak element ratios.

Figure~\ref{f:OFe} shows the evolution of the [O/Fe] ratio as a
function of metallicity. All these ratios are recomputed assuming
[O/H]$_{\sun}=-3.26$ (Holweger 2001). In most cases, this ratio is is
sub-solar at high metallicities ([Fe/H]$>$-1.6). However, [O/Fe] is
still greater than solar at low metallicities. Similarly, Galactic
metal-poor stars show an enhancement between 0.35 and 1 (Goswami \&
Prantzos 2000 and references therein). However, in the case of high
[O/Fe] ratio, one cannot decipher whether dust depletion or
nucleosynthesis produces such high values. In the case of low [O/Fe],
however, we know that the intrinsic values cannot be much smaller than
the values observed.

Models from Hou, Boissier \& Prantzos (2001) seem to underproduce the
corresponding ratios with respect to the observations for both DLAs
and sub-DLAs. In contrast, models from Calura, Matteucci \&
Vladilo (2002) seem able to reproduce the
observed sub-solar values of [O/Fe] (see also Chiappini, Romano \&
Matteucci 2002). At present, however, the observed sample is still too 
small to allow firm conclusions to be drawn.

\begin{figure*}
\psfig{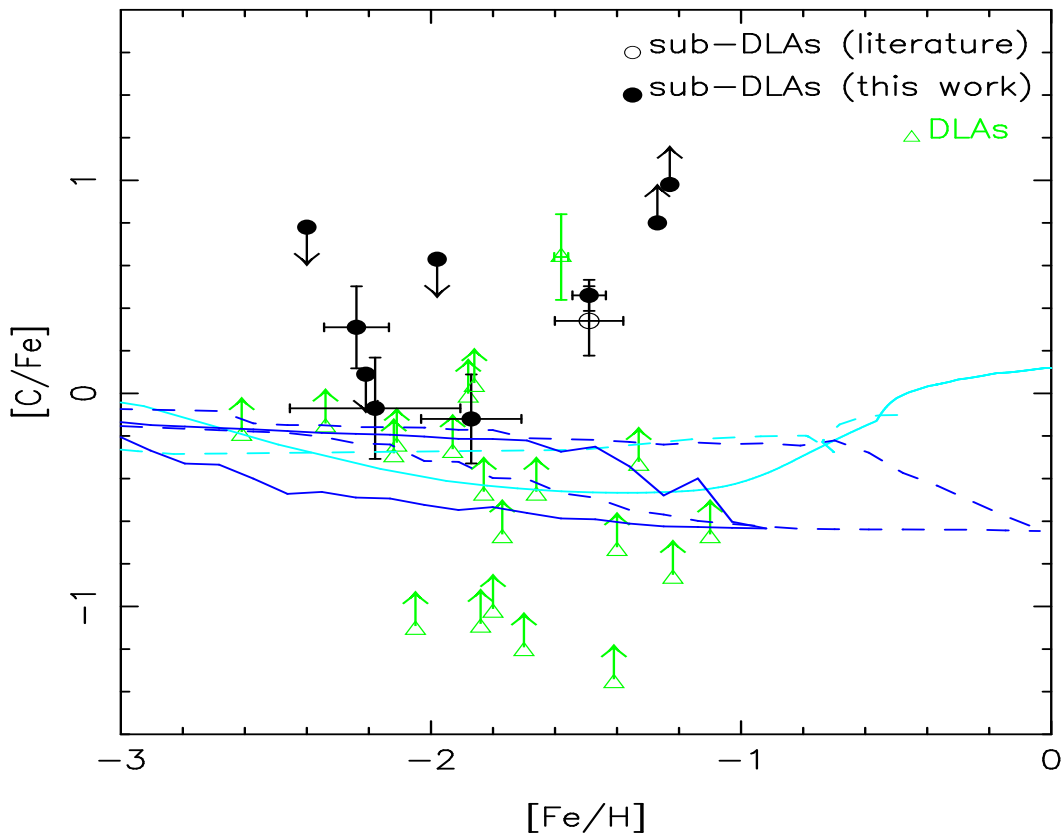}
\caption{[C/Fe] ratio as a function of metallicity. Sub-DLAs are 
markedly more over-solar than DLAs. The dark-coloured solid (dashed)
curves are contours for sub-DLAs (DLAs) column densities derived from
models of disc galaxies (Hou, Boissier \& Prantzos 2001). The
light-coloured solid (dashed) curves are for spirals (irregulars)
representative of sub-DLAs (DLAs) in the models of (Calura, Matteucci
\& Vladilo 2003).}
\label{f:CFe}
\end{figure*}

\begin{figure*}
\psfig{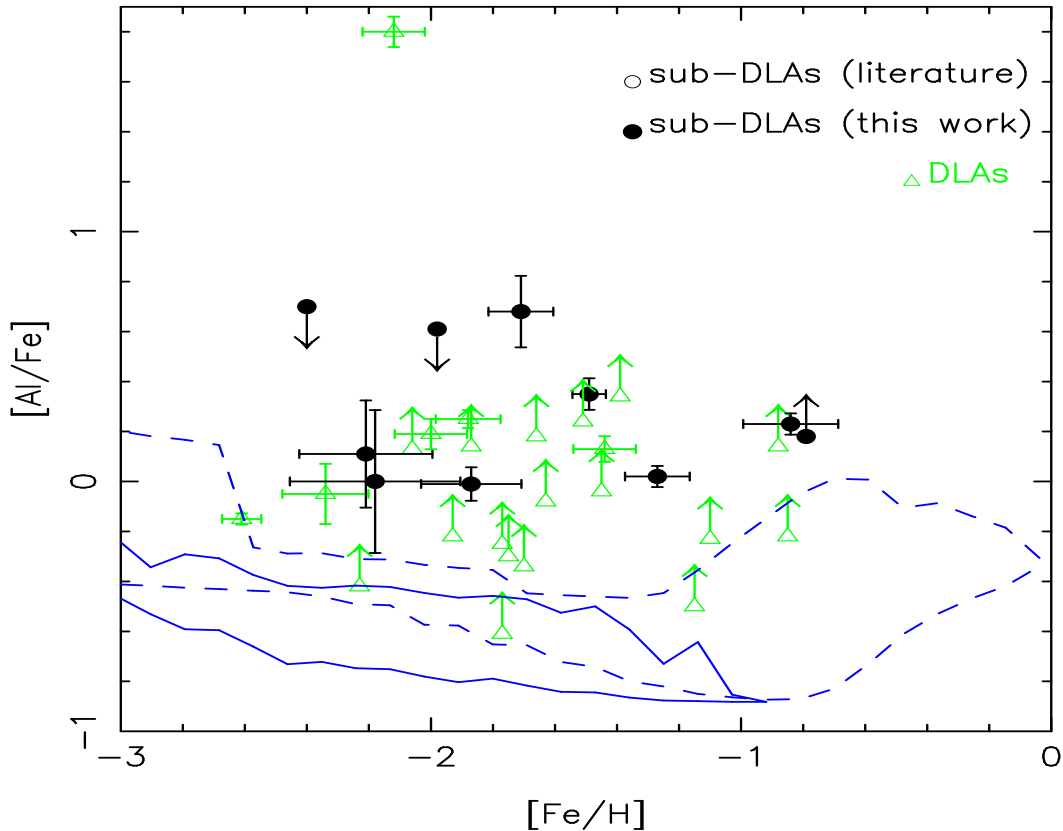}
\caption{[Al/Fe] ratio as a function of metallicity. Again, sub-DLAs are 
remarkably more over-solar than DLAs. The solid (dashed) curves are
contours for sub-DLAs (DLAs) column densities derived from models of
disc galaxies (Hou, Boissier \& Prantzos 2001).
}
\label{f:AlFe}
\end{figure*}

{\bf [C/Fe]:} For the same reasons as O~{\sc i}, the first
comprehensive set of measurements of C~{\sc ii} has been made in high
HI absorbers. [C/Fe] is approximately solar in the majority of the
Galactic stars. Previous studies of DLAs only allowed for lower limit
determinations (see Figure~\ref{f:CFe}), but measurements of C~{\sc
ii} in sub-DLAs provide a tracer of Fe-peak elements regardless of
dust depletion. In this figure we use [C/H]$_{\sun}=-3.41$ (Holweger
2001) as an estimate of the solar value.  Some of the [C/Fe] ratios in
sub-DLAs are clearly oversolar, at the level of the overabundance
observed in the two tentative [C/Fe] measurements in the DLAs. Since C
is a mildly refracted element in the interstellar medium, these
measurements provide our best indicator of the dust content of
sub-DLAs. Indeed, although C and Fe do not share the same
nucleosynthesis origin, in the Milky Way stars, C and Fe show
identical behaviours, i.e. [C/Fe]=0 (Goswami \& Prantzos
2000). Therefore the deviation from zero observed in sub-DLAs are
suggestive of either the presence of dust in these systems or a
different behaviour of C relative to Fe relative to the Milky Way
stars. Both types of chemical evolution models predict abundance
ratios in the intermediate range between DLAs and
sub-DLAs. Nevertheless, the models of Hou, Boissier \& Prantzos (2001)
do not include yields from intermediate-mass stars. This will affect
the [C/Fe] abundance ratio. The models of Calura, Matteucci \& Vladilo
(2002) predict very little difference between DLAs and sub-DLAs, as is
observed. But again, given the limited data set, these conclusions are
tentative.

{\bf [Al/Fe]:} The metallicity evolution of the [Al/Fe] ratio is
presented in Figure~\ref{f:AlFe}, using [Al/H]$_{\sun}=-5.51$
(Grevesse \& Sauval 1998). In DLAs, this ratio follows a trend similar
to the one observed in metal-poor stars (Prochaska \& Wolfe 2002). In
the sub-DLA case, this ratio is strongly affected by ionization
corrections which increase the real abundance of Al
with respect to that measured solely from Al~{\sc ii}. The [Al/Fe] ratio
will consequently rise by a factor of 0.3 dex in the worst case (see
Paper I). Again sub-DLAs are characterised by a mean [Al/Fe] elemental
ratio larger than DLAs. In this case, we note that models of Hou,
Boissier \& Prantzos (2001) do not match the values of high-redshift
DLAs nor the observations in the Milky Way. It should be noted that
the nucleosynthesis situation of this element is unclear (Goswami \&
Prantzos 2000) but it could also be that the intermediate-mass stars
not included in the models contribute to the true [Al/Fe] abundance
ratio.

\section{Conclusion}

P\'eroux \e\ (2003) have extended the DLA definition to a new class of
absorbers: the sub-DLAs with column density $10^{19}$ $<$ N(HI) $<$ 2
$\times 10^{20}$ atoms cm$^{-2}$. These systems are believed to
contain a large fraction of the neutral gas mass in the Universe, 
especially at
$z>3.5$. Based on these considerations, we have constructed and fully
analysed a sample of 12 sub-DLAs (see Paper I). In the present paper,
we analyse several of the properties of these absorbers in conjunction
with DLAs from the literature. Our main findings can be summarised as
follows:

\begin{itemize}

\item Our sample of sub-DLAs can be used to observationally determine for 
the first time the shape of the column density distribution, $f(N)$,
down to N(HI) $=10^{19}$ cm$^{-2}$. The results are in good agreement
with the predictions from P\'eroux \e\ (2003). An evolutionary study
of $f(N)$ is not possible with the current sample illustrating the
need for a sample of sub-DLAs at high-redshifts.

\item An analysis of the clustering of sub-DLAs in our sample 
shows that these are more clustered than expected from a random
distribution, although the number statistic is too small to draw firm
conclusions.

\item We measure the kinematic properties of the sub-DLAs from our sample 
together with a statistically significant number of DLAs from the
literature. We compare low- and high-ionization transition widths and
find that the sub-DLA properties roughly span the parameter space of
the DLAs, thus challenging multicomponent semi-analytical models, such
as the one presented by Maller \e\ (2002).

\item The metallicity of absorbers as traced by [Fe/H] shows a slightly more 
pronounced slope for sub-DLAs ($\alpha=-0.40\pm0.22$) than for DLAs
($\alpha=-0.18\pm0.12$). In addition, the HI-weighted mean metallicity
is computed for various sub-sets of quasar absorbers. The evolution of
$[\langle \rm Fe/\rm H_{\rm DLA}\rangle]$ might be stronger for
sub-DLAs than for DLAs, and absorbers with N(HI) $>10^{21.0}$ atoms
cm$^{-2}$ appear to be the less evolved, suggesting these objects in
particular should be studied in detail. Observational evidence
supports the hypothesis that this different behaviour is {\it not} due
to the hidden effect of dust. We therefore propose that sub-DLAs might
be associated with a class of galaxies which better traces the overall
chemical evolution of the Universe.

\item A study of the metallicity evolution with metal line profile ionization
width might show hints of a correlation, whereby higher [Fe/H] ratios
are associated with systems with larger widths. However the
statistical significance of this result is low and a larger sample of
observations is required to unambiguously address this issue.

\item Abundance ratios for [Si/Fe], [O/Fe], [C/Fe] and [Al/Fe] were
determined and compared with two different sets of models of the
chemical evolution of galaxies. Overall, these appear to resemble
abundance ratios observed in DLAs. The question thus remains open
whether the two classes of objects have similar chemical evolution
histories or whether there are objects with different ages and star
formation histories, but at a similar stage of chemical evolution. The
first comprehensive sets of measurements of O~{\sc i} and C~{\sc ii}
in high H~{\sc i} column density systems are given. Indeed, these are
well-defined in sub-DLAs while they are almost always saturated in
DLAs. These elements, unaffected by dust depletion, provide direct
indicators of the abundances in these systems.

\end{itemize}

These various issues illustrate the importance of further studies of
sub-DLAs to interpret the overall chemical evolution
of neutral matter with redshift, as well as to extend the analysis of
quasar absorbers properties to a lower column density range. We are
currently undertaking an observational program aimed at studying
sub-DLAs at high-redshifts.

\section{Acknowledgments}
We are grateful to Samuel Boissier for extending the predictions of
Hou, Boissier \& Prantzos (2002) to the sub-DLA column densities and
to Francesco Calura for doing the same with the models of Calura,
Matteucci \& Vladilo (2002). We have benefited from conversations with
Samuel Boissier, Francesco Calura, Mike Irwin, Ari Maller, Francesca
Matteucci, Paolo Molaro, Nikos Prantzos, Simone Recchi and Zheng
Zheng. We also thank an anonymous referee for extensive and helpful
comments. This work was supported in part by the European Community
Research and Training Network ``The Physics of the Intergalactic
Medium''. CP is founded by an European Marie Curie Fellowship and MDZ
by Swiss National Funds. RGM thanks the Royal Society.

\bsp \label{lastpage} 
\end{document}